\documentclass[leqno,11pt]{amsart}
\usepackage{amsmath,amsfonts,amsthm,amssymb,indentfirst,epic,url,centernot}
\usepackage{color}
\usepackage{graphicx}
\usepackage{caption}
\usepackage[utopia]{mathdesign}
\usepackage{algorithm}
\usepackage[noend]{algorithmic} 
\algsetup{indent=3mm} 

\newtheorem{Theorem}{Theorem}
\newtheorem{Example}{Example}[section]
\newtheorem{Remark}{Remark}
\newtheorem{Notation}{Notation}

\newtheorem{Definition}{Definition}[section]
\newtheorem{Lemma}{Lemma}

\newtheorem{Proposition}{Proposition}
\newtheorem{Corollary}{Corollary}
\newenvironment{Proof}[1][Proof]{\noindent {\bf Proof.}}{\qed}

\begin{document}

\title{On the Lattice of Cyclic Linear Codes Over Finite Chain Rings}



\author{Alexandre Fotue Tabue}
\address{Department of mathematics, Faculty of Sciences,  University of Yaoundé 1, Cameroon} \email{alexfotue@gmail.com}
\author{Christophe Mouaha}
\address{Department of mathematics,  Higher Teachers Training College of Yaoundé, University of Yaoundé 1, Cameroon} \email{cmouaha@yahoo.fr}

\begin{abstract}   Let $\texttt{R}$ be a commutative finite chain ring of invariants $(q,s).$ In this paper, the trace representation of any
free cyclic $\texttt{R}$-linear code of length $\ell,$ is
presented, via the $q$-cyclotomic cosets modulo $\ell,$ when
$\texttt{gcd}(\ell,q)=1.$ The lattice $\left\langle\,\texttt{Cy}
(\texttt{R},\ell);+,\cap\right\rangle$ of cyclic
$\texttt{R}$-linear codes of length $\ell,$ is investigated. A
lower bound on the Hamming distance of cyclic $\texttt{R}$-linear
codes of length $\ell$, is established. When $q$ is even, a family
of MDS and self-orthogonal $\texttt{R}$-linear cyclic codes, is
constructed.
\end{abstract}

\maketitle

\section{Introduction}

Research on linear codes over chain rings can be found in
\cite{HL00,NS00} and cyclic linear codes were among the first
codes practically used and they play a very significant role in
coding theory. For instance, cyclic codes can be efficiently
encoded using shift registers. Many important codes such as the
Golay codes, Hamming codes and BCH codes can be represented as
cyclic codes. Cyclic linear codes have been studied for decades
and a lot of progress has been made (see, for example,
\cite{WS77}).

 Let $\texttt{R}$ be finite chain ring with invariant $(q,s),$ and
$\ell$ a positive integer such that $\texttt{gcd}(q,\ell)=1.$ An
$\texttt{R}$-\emph{linear code} of length $\ell$ is an
$\texttt{R}$-submodule of $\texttt{R}^\ell.$  An
$\texttt{R}$-linear code $\mathcal{C}$ of length $\ell$ is
\index{cyclic operator}\emph{cyclic} if for every codeword
$(\textbf{c}_0, \textbf{c}_1, \cdots , \textbf{c}_{\ell-1})$ in
$\mathcal{C}$ then the word $(\textbf{c}_{\ell-1}, \textbf{c}_0,
\cdots , \textbf{c}_{\ell-2})$ also belongs to $\mathcal{C}.$ For
example, the zero module $\{\underline{\textbf{0}}\},$ the trivial
$\texttt{R}$-linear code $\texttt{R}^\ell$ and the repetition code
$\textbf{1}:=\{(\textbf{c},\textbf{c},\cdots,\textbf{c})\;:\;\textbf{c}\in\texttt{R}\}$
are cyclic $\texttt{R}$-linear codes of length $\ell.$

As usual, the $\texttt{R}$-module isomorphism
\begin{align}\begin{array}{cccc}
 \Psi: & \texttt{R}^\ell & \rightarrow & \texttt{R}[X]/ (X^\ell-1) \\
       & (\textbf{c}_0,\textbf{c}_1,\cdots,\textbf{c}_{\ell-1}) & \mapsto &
       \textbf{c}_0+\textbf{c}_1X+\cdots+\textbf{c}_{\ell-1}X^{\ell-1}+(X^\ell-1)
  \end{array}\end{align} is used to identify the cyclic $\texttt{R}$-linear codes of length
$\ell$ with ideals of the ring $\texttt{R}[X]/(X^\ell-1)$ and
$(X^\ell-1)$ is the ideal of $\texttt{R}[X]$ generated by
$X^\ell-1.$ The $\texttt{R}$-module $\Psi(\mathcal{C})$ is called
\index{polynomial representation}\emph{polynomial representation}
of the $\texttt{R}$-linear code $\mathcal{C}.$ This polynomial
representation of cyclic linear codes is used by Calderbank and
Sloane in \cite{CS95}, by Kanwar and López-Permouth \cite{KL97},
for study the structure of cyclic linear codes over
$\mathbb{Z}_{p^s}.$ Wan \cite{Wan99} extended these results to
cyclic linear codes over Galois rings. Norton and Sãlãgean
\cite{NS00} and Dinh and López-Permouth \cite{DP04}, in turn,
extended the results of \cite{CS95} and \cite{KL97} to cyclic
linear codes over finite chain rings.

 We denote by $\texttt{Cy}(\texttt{R},\ell),$ the set of all cyclic
$\texttt{R}$-linear codes of length $\ell.$ We equip
$\texttt{Cy}(\texttt{R},\ell)$ of binary operations as: $+,$ and
$\cap.$ When $s=1$ and $\texttt{gcd}(q,\ell)=1,$ it is widely
known that $\mathbb{F}_q[X]$ is a domaine ideal ring. Hence the
lattice of ideals of $\mathbb{F}_q[X]/(X^\ell-1)$ is distributive.
By the polynomial representation, it follows that for every
$\mathcal{C}\in\texttt{Cy}(\mathbb{F}_q,\ell),$ there exists a
unique monic divisor $g$ of $X^\ell-1$ in $\mathbb{F}_q[X]$ such
that $\Psi(\mathcal{C})= \langle g\rangle,$ where  $\langle
g\rangle:=(g)/(X^\ell-1),$ and $g$ is called the \emph{generator
polynomial} of $\mathcal{C},$ and
\begin{align}\Psi^{-1}(\langle g_1\rangle)\cap\Psi^{-1}(\langle
g_2\rangle)=\Psi^{-1}(\langle \texttt{lcm}(g_1,g_2)\rangle)\text{
and }\Psi^{-1}(\langle g_1\rangle)+\Psi^{-1}(\langle
g_2\rangle)=\Psi^{-1}(\langle\texttt{gcd}(g_1,g_2)\rangle),\end{align}
for all monic divisors $g_1$ and $g_2$ of $X^\ell-1$ in
$\mathbb{F}_q[X].$ Hence, the lattice
$\left\langle\texttt{Cy}(\mathbb{F}_q,\ell);\;+,\cap;\{\underline{\textbf{0}}\},\mathbb{F}_q^\ell\right\rangle$
is distributive. In the general case $s\neq 1,$ the lattice
$\left\langle\texttt{Cy}(\texttt{R},\ell);\;+,\cap\right\rangle$
is little known. In  \cite{NS00}, Norton and Sãlãgean, describe
the generating set in standard form of any cyclic
$\texttt{R}$-linear code of length $\ell.$ The objective of this
paper is to develop another approach of construction of cyclic
linear codes in case the length of the code is coprime to the
characteristic of the finite chain ring. Fundamental theory of
this approach will be developed, and will be employed to construct
any cyclic linear code. BCH-bound of any cyclic linear code will
be also established.

The paper is organized as follows: Some definitions and results
about the finite chain rings and Galois extension of finite chain
rings, are recalled in Section \ref{Sect:2}. Section \ref{Sect:3}
discusses the notions of the type of any linear code over a finite
chain ring. In Section \ref{Sect:4} the set of cyclotomic
partitions is investigated. In Section \ref{Sect:5}, the
trace-description of free cyclic linear codes over finite chain
rings is presented. A lattice of cyclic linear codes is studied in
Section \ref{Sect:6}.

\section{Background on finite chain rings}\label{Sect:2}

Throughout of this section, $\texttt{R}$ is a finite local ring
with identity, $\texttt{J}(\texttt{R})$ denotes the maximal ideal
of $\texttt{R},$ and $\mathcal{M}_{u\times\ell}(\texttt{R})$
denotes the set of all $u\times\ell$-matrices over $\texttt{R}$
for all $1\leq u\leq\ell.$

\begin{Definition}
A finite local ring $\texttt{R}$ with identity is called a
\emph{finite chain ring of invariants $(q,s)$} if
\begin{enumerate}
    \item $\texttt{J}(\texttt{R})$ is principal,
$\texttt{R}/\texttt{J}(\texttt{R})\simeq\mathbb{F}_q,$ and;
    \item $\texttt{R}\supsetneq \texttt{R}\theta \supsetneq \cdots\supsetneq
\texttt{R}\theta^{s-1} \supsetneq \texttt{R}\theta^s
=\{0_\texttt{R}\}$ where $\theta$ is a generator of
$\texttt{J}(\texttt{R}).$
\end{enumerate}
\end{Definition}

For example, for every positive integer $s,n, p$ and $p$ prime,
the ring $\mathbb{Z}_{p^s},$ is a finite chain ring of invariants
$(p,s),$ and the ring $\mathbb{F}_{p^n}[\theta]$ with
$\theta^s=0_\texttt{R}$ and $\theta^{s-1}\neq 0_\texttt{R},$ is a
finite chain ring of invariants $(p^n,s).$

 Let $\texttt{R}$ be a finite chain
ring of invariants $(q,s)$ and $\theta$ be a generator of
$\texttt{J}(\texttt{R}).$ The ring epimorphism
$$\begin{array}{cccc}
                               \pi: & \texttt{R} & \rightarrow & \mathbb{F}_q \\
                                   & \textbf{c} & \mapsto &
                                   \textbf{c}\,(\texttt{mod}\;\theta)
                               \end{array}$$
naturally extends to $\texttt{R}[X],$ of the following way:
$\pi(\sum \textbf{a}_iX^i)=\sum \pi(\textbf{a}_i)X^i$ and
$\mathcal{M}_{u\times\ell}(\texttt{R})$ of the following way:
$\pi$ acts on all the coefficients of any polynomial (resp.
matrix) over $\texttt{R}.$ Let $\texttt{R}^\times$ be the
multiplicative group of units of $\texttt{R}.$ Obviously, the
cardinality of $\texttt{R}^\times$ is $q^{(s-1)}(q-1).$ It follows
that
$\texttt{R}^\times\simeq\Gamma(\texttt{R})^*\times(1+\texttt{J}(\texttt{R}))$
with $\Gamma(\texttt{R})^*=\{ \textbf{b}\in \texttt{R}\,:\,
\textbf{b}\neq 0_\texttt{R},\, \textbf{b}^{q}=\textbf{b} \},$ and
$\Gamma(\texttt{R})^*$ is a cyclic subgroup of
$\texttt{R}^\times,$ of order $q-1.$ The set
$\Gamma(\texttt{R})=\Gamma(\texttt{R})^*\cup\{0_\texttt{R}\}$ is
called the \emph{Teichmüller set} of $\texttt{R}.$

We say that the ring $\texttt{S}$ is an \emph{extension} of
$\texttt{R}$ and we denote it by $\texttt{S}|\texttt{R}$ if
$\texttt{R}$ a subring of $\texttt{S}$ and $1_\texttt{R} =
1_\texttt{S}.$ We denote by
$\texttt{rank}_\texttt{R}(\texttt{S}),$ the rank of
$\texttt{R}$-module $\texttt{S}.$ Let $f\in\texttt{R}[X]$ of
degree $m$ and $(f)$ is an ideal of $\texttt{R}[X]$ generated by
$f.$  Let $f$ be a monic polynomial over $\texttt{R}$ of degree
$m.$ We say that $f$ is \emph{basic irreducible}(resp. basic
primitive ) if $\pi(f)$ is irreducible over $\mathbb{F}_q$ (resp.
primitive).

\begin{Definition}\label{D-Gal} Let  $\texttt{R}$ be a finite chain ring of invariants $(q,s).$
A finite ring $\texttt{S}$ is a \emph{Galois extension} of
$\texttt{R}$ of degree $m,$ if
$\texttt{S}\simeq\texttt{R}[\alpha],$ (as $\texttt{R}$-algebras)
where $\alpha$ is a root of a monic basic primitive polynomial
over $\texttt{R}$ of degree $m.$
\end{Definition}

\begin{Remark} Let  $\texttt{R}$ be a finite chain ring of invariants
$(q,s)$ and $\texttt{S}$ is a  Galois extension  of $\texttt{R}$
of degree $m.$ Let $\xi$ be a generator of
$\Gamma(\texttt{S})\setminus\{0_\texttt{S}\}.$ Then $\texttt{S}$
is also a finite chain ring of invariants $(q^m,s)$ and
$\texttt{S}=\texttt{R}[\xi].$
\end{Remark}

We denote by $\texttt{Aut}_\texttt{R}(\texttt{S}),$ the group of
ring automorphisms of $\texttt{S}$ which fix the elements of
$\texttt{R}.$

\begin{Theorem}\label{cGal}\cite[Chap III.; Proposition 2.1(1)]{DI71} Let  $\texttt{R}$ be a finite chain ring of invariants
$(q,s).$ The  ring $\texttt{S}$ is a  Galois extension  of
$\texttt{R}$ of degree $m$ if and only if
$\texttt{R}=\{\textbf{a}\in\texttt{S}\;:\;\sigma(\textbf{a})=\textbf{a}\;\text{
for all } \sigma\in \texttt{Aut}_\texttt{R}(\texttt{S})\}$ and
$\texttt{J}(\texttt{S})=\texttt{S}\texttt{J}(\texttt{R}).$
\end{Theorem}

For example, for positive integers $p,n,s$ and $p$ prime, the
Galois extension of $\mathbb{Z}_{p^s}$ of degree $n,$ is the
Galois ring $\texttt{GR}(p^s,n)\simeq\mathbb{Z}_{p^s}[X]/(f),$
where $f$ is a monic basic irreducible over $\texttt{R},$ of
degree $n.$

\begin{Proposition}\cite[Theorem XV.10]{McD74} Let $\texttt{R}$ be a finite chain ring of invariants $(q,s)$
and $\texttt{S}$ be the Galois extension of $\texttt{R}$ of degree
$m.$ Let $\xi$ be a generator of
$\Gamma(\texttt{S})\setminus\{0_\texttt{S}\}.$ Then
\begin{enumerate}
    \item $\texttt{S}$ is a free $\texttt{R}$-module with free $\texttt{R}$-basis $\{1,\xi,\cdots,\xi^{m-1}\};$
   \item $\texttt{Aut}_\texttt{R}(\texttt{S})$ is a cyclic group of order
   $m,$ and a generator of $\texttt{Aut}_\texttt{R}(\texttt{S})$
   is given by $\sigma:\xi\mapsto\xi^q.$
\end{enumerate}
\end{Proposition}

\begin{Definition} Let $\texttt{S}|\texttt{R}$ be the Galois extension of finite
chain rings of degree $m$ and $\sigma$ be a generator of
$\texttt{Aut}_\texttt{R}(\texttt{S}).$ The map
$\texttt{Tr}_\texttt{R}^\texttt{S}:=\sum\limits_{i=0}^{m-1}\sigma^i,$
is called the \emph{trace map} of $\texttt{S}|\texttt{R}.$
\end{Definition}

\begin{Theorem}\label{epi-tr}\cite[Chap III.;Corollary 2.2]{DI71} Let $\texttt{S}|\texttt{R}$ be the Galois extension
of finite chain rings. Then $\texttt{Tr}_\texttt{R}^\texttt{S}$ is
a generator of $\texttt{S}$-module
$\texttt{Hom}_\texttt{R}(\texttt{S},\texttt{R}).$
\end{Theorem}

\begin{Proposition}\label{Gsr} Let $\texttt{R}$ be a finite chain ring of invariants $(q,s)$
and $\texttt{S}$ be the Galois extension of $\texttt{R}$ of degree
$m.$ Then  for every positive integer $z,$ for every generator
$\xi$\, of\, $\Gamma(\texttt{S}),$ the ring $\texttt{R}[\xi^{zq}]$
is the Galois extension of $\texttt{R}$ of degree
    $m_z,$ where $m_z:=\texttt{min}\{i\in\mathbb{N}\setminus\{0\}\;:\;zq^{i}\equiv
1\;(\texttt{mod}\;(q^m-1))\}.$
\end{Proposition}

\begin{Proof} We set
$f:=(X-\xi^{zq})(X-\xi^{zq^2})\cdots(X-\xi^{zq^{m_z}}).$ Since
$\texttt{S}|\texttt{R}$ is Galois extension, we deduce by
Theorem\,\ref{cGal}, that $f\in\texttt{R}[X]$ and $\pi(f)$ is
irreducible. Hence $f$ is a basic irreducible polynomial over
$\texttt{R}$ and the degree of $f$ is $m_z.$ It follows that
$\texttt{R}[\xi^{zq}]$ is a Galois extension of $\texttt{R}$ of
degree $m_z,$ because by Definition\,\ref{D-Gal}, $\xi^{zq}$ is a
root of a basic irreducible polynomial over $\texttt{R}$ of degree
of $m_z.$
\end{Proof}

\section{Linear codes over finite chain rings}\label{Sect:3}

For this section, $\texttt{R}$ is a finite chain ring of
invariants $(q,s),$ and $\theta$ is a generator of maximal ideal
$\texttt{J}(\texttt{R}).$ The ring epimorphism
$\pi:\texttt{R}\rightarrow\mathbb{F}_q$ naturally extends to
$\texttt{R}^\ell$ of the following way:
$\pi(\underline{\textbf{c}}):=(\pi(\textbf{c}_0),\pi(\textbf{c}_1),\cdots,\pi(\textbf{c}_{\ell-1})),$
for every
$\underline{\textbf{c}}:=(\textbf{c}_0,\textbf{c}_1,\cdots,\textbf{c}_{\ell-1})\in\texttt{R}^\ell.$
Recall that an $\texttt{R}$-linear code of length $\ell$ is an
$\texttt{R}$-submodule of $\texttt{R}^\ell.$ We say that an
$\texttt{R}$-linear code is \index{free linear code}\emph{free} if
it is free as $\texttt{R}$-module.

\subsection{Type and rank of a linear code}

  An $k\times\ell$-matrix $G$ over $\texttt{R},$ is called a \index{generator matrix}\emph{generator matrix} for $\mathcal{C}$
if the rows of $G$ span $\mathcal{C}$ and none of them can be
written as an $\texttt{R}$-linear combination of other rows of
$G.$ We say that $G$ is a generator matrix in
\index{standard form}\emph{standard form} if \begin{align}\label{stG}G=\left(%
\begin{array}{cccccc}
  I_{k_{0}}  & G_{0,1}        & G_{0,2}        &\cdots    & G_{0,s-1}                 & G_{0,s}                \\
  0        &  \theta I_{k_{1}} &   \theta G_{1,2} & \cdots   &  \theta G_{1,s-1}          & \theta  G_{1,s}         \\
  \cdots   & \cdots         & \cdots         & \cdots   &  \cdots                   & \cdots               \\
  0        &      0         &           0    & \cdots        & \theta^{s-1} I_{k_{s-1}}  &  \theta^{s-1} G_{s-1,s}
\end{array}%
\right)U,
\end{align}
where $U$ is a suitable permutation matrix and
$G_{t,j}\in\mathcal{M}_{k_t\times k_j}(\texttt{R}).$ The $s$-tuple
$(k_{0}, k_{1},\cdots,k_{s-1})$ is called \index{type of a
matrix}\emph{type} of $G$ and
$\texttt{rank}(G)=k_0+k_1+\cdots+k_{s-1}$ is the \index{rank of a
matrix}\emph{rank} of $G.$

\begin{Proposition}(\cite[Proposition 3.2, Theorem 3.5]{NS00}) Each $\texttt{R}$-linear
code $\mathcal{C}$ admits a generator matrix $\texttt{G}$ standard
form. Moreover, the type is the same for any generator matrix in
standard form for $\mathcal{C}.$
\end{Proposition}

\begin{Definition} Let $\mathcal{C}$ be an $\texttt{R}$-linear
code and $G$ be a generator matrix $\texttt{G}$ standard form. The
rows of $G$ form an $\texttt{R}$-basis for $\mathcal{C},$
so-called \emph{standard} $\texttt{R}$-basis for $\mathcal{C}.$
\end{Definition}

So the type and the rank are the invariants of $\mathcal{C},$ and
henceforth we have the following definition.

\begin{Definition} Let $\mathcal{C}$ be an $\texttt{R}$-linear code.
\begin{enumerate}
    \item The \index{type of a linear code}\emph{type} of $\mathcal{C}$ is the type of a generator matrix of $\mathcal{C}$ in standard form.
    \item The \index{rank of a linear code}\emph{rank} of $\mathcal{C},$ denoted $\texttt{rank}_\texttt{R}(\mathcal{C}),$  is the rank of a generator matrix of $\mathcal{C}$ in standard form.
\end{enumerate}
\end{Definition}

Obviously, any $\texttt{R}$-linear code $\mathcal{C}$ of length
$\ell$ of type $(k_{0}, k_{1},\cdots,k_{s-1})$ is free if and only
if the rank of $\mathcal{C}$ is $k_0,$ and
$k_1=k_2=\cdots=k_{s-1}=0.$ It defines the scalar product on
$\texttt{R}^\ell$ by:
$\underline{\textbf{a}}\cdot\underline{\textbf{b}}^{\texttt{T}}:=\sum\limits_{i=0}^{\ell-1}\textbf{a}_i\textbf{b}_i,$
where $\underline{\textbf{b}}^{\texttt{T}}$ is the transpose of
$\underline{\textbf{b}}.$ Let $\mathcal{C}$ be an
$\texttt{R}$-linear code of length $\ell.$ The dual code of
$\mathcal{C},$ denoted $\mathcal{C}^\perp,$ is an
$\texttt{R}$-linear code of length $\ell,$ is defined by:
$\mathcal{C}^\perp:=\left\{\underline{\textbf{a}}\in\texttt{R}^\ell\;:\;\underline{\textbf{a}}\cdot\underline{\textbf{b}}^{\texttt{T}}=0_\texttt{R}\text{
for all }\underline{\textbf{b}}\in\mathcal{C}\right\}.$ A
generator matrix of $\mathcal{C}^\perp,$ is called parity-check
matrix of $\mathcal{C}.$ We recall that an $\texttt{R}$-linear
code $\mathcal{C}$ is \emph{self-orthogonal} if
$\mathcal{C}\subseteq\mathcal{C}^{\perp}.$ An $\texttt{R}$-linear
code $\mathcal{C}$ is \emph{self-dual} if
$\mathcal{C}=\mathcal{C}^{\perp}.$

\begin{Proposition}\label{dual-ope}(\cite[Theorem 3.1]{HL00})
Let $\mathcal{C}$ and $\mathcal{C}'$ be $\texttt{R}$-linear codes
of length $\ell.$  Then
$(\mathcal{C}+\mathcal{C}')^\perp=\mathcal{C}^\perp\cap\mathcal{C}'^\perp,$
$(\mathcal{C}\cap\mathcal{C}')^\perp=\mathcal{C}^\perp+\mathcal{C}'^\perp,$
and $(\mathcal{C}^\perp)^\perp=\mathcal{C}.$
\end{Proposition}

\begin{Proposition}\label{dual-type}(\cite[Theorem 3.10]{NS00})
Let $\mathcal{C}$ be an $\texttt{R}$-linear code of length $\ell,$
of type $(k_{0}, k_{1},\cdots,k_{s-1}).$  Then
\begin{enumerate}
    \item the type of $\mathcal{C}^\perp$ is $(\ell-k,
k_{s-1},\cdots,k_{1}),$ where $k:=k_{0}+k_{1}+\cdots+k_{s-1}.$
    \item $|\mathcal{C}|=q^{\sum\limits_{t=0}^{s-1}(s-t)k_t},$
    where $|\mathcal{C}|$ denotes the number of elements of $\mathcal{C}.$
\end{enumerate}
\end{Proposition}

\begin{Definition}\label{D-ann} Let $\texttt{R}$ be a finite chain
ring of invariants $(q,s)$ and $\theta$ be a generator of
$\texttt{J}(\texttt{R}).$ Let $\mathcal{C}$ be an
$\texttt{R}$-linear code of rank $k.$
\begin{enumerate}
    \item The $\texttt{R}$-linear subcode
$\texttt{Annih}_{\mathcal{C}}(\theta):=\left\{\underline{\textbf{c}}\in\mathcal{C}\;:\;\theta\underline{\textbf{c}}=\underline{\textbf{0}}\right\},$
of $\mathcal{C},$ is called the \emph{annihilator} of
$\mathcal{C}.$
    \item  The \emph{residue code} of $\mathcal{C}$ is the $\mathbb{F}_q$-linear code
$\pi(\mathcal{C}):=\left\{\pi(\underline{\textbf{c}})\;:\;\underline{\textbf{c}}\in\mathcal{C}\right\}.$
\end{enumerate}
\end{Definition}

\begin{Remark}\label{Rquot} Let $\mathcal{C}$ be an $\texttt{R}$-linear code with generator matrix $G,$ as in (\ref{stG}). Then a generator
matrix of $\texttt{Annih}_{\mathcal{C}}(\theta)$ is $$\theta^{s-1}\left(%
\begin{array}{cccccc}
  I_{k_{0}}  & G_{0,1}        & G_{0,2}        &\cdots    & G_{0,s-1}                 & G_{0,s}                \\
  0        &    I_{k_{1}} &     G_{1,2} & \cdots   &    G_{1,s-1}          &    G_{1,s}         \\
  \cdots   & \cdots         & \cdots         & \cdots   &  \cdots                   & \cdots               \\
  0        &      0         &           0    & \cdots        &  I_{k_{s-1}}  &   G_{s-1,s}
\end{array}%
\right)U.$$
\end{Remark}

The \emph{Hamming distance} of an $\texttt{R}$-linear code
$\mathcal{C}$ of length $\ell,$  is defined as:
$$\texttt{wt}(\mathcal{C}):=\texttt{min}\left\{\texttt{wt}(\underline{\textbf{c}})\,:\,
\underline{\textbf{c}}\in\mathcal{C}, \underline{\textbf{c}}\neq
\underline{\textbf{0}}\right\},$$ where
$\texttt{wt}(\textbf{c}_0,\textbf{c}_1,\cdots,\textbf{c}_{\ell-1}):=|\{j\in\Sigma_\ell\;:\;\textbf{c}_j\neq
0_\texttt{R}\}|.$

\begin{Theorem}\label{dis} Let $\mathcal{C}$ be an
$\texttt{R}$-linear code. Then
$\texttt{wt}(\mathcal{C})=\texttt{wt}(\texttt{Annih}_{\mathcal{C}}(\theta)).$
\end{Theorem}

\begin{Proof} We set $\mathcal{A}:=\texttt{Annih}_{\mathcal{C}}(\theta).$
Let $\underline{\textbf{c}}$ in $\mathcal{C}$ with
$\texttt{wt}(\underline{\textbf{c}}) = \texttt{wt}(\mathcal{C}).$
Let $t$ be the least positive integer such that
$\theta^t\underline{\textbf{c}}=\underline{\textbf{0}}.$ Then
$t\neq 0$ and
$\theta^{t-1}\underline{\textbf{c}}\in\mathcal{A}\backslash\{\underline{\textbf{0}}\}.$
Hence we must have: $\texttt{wt}(\mathcal{A})\leq
\texttt{wt}(\theta^{t-1}\underline{\textbf{c}})\leq
\texttt{wt}(\underline{\textbf{c}})=\texttt{wt}(\mathcal{A}).$
Hence the equality
$\texttt{wt}(\mathcal{C})=\texttt{wt}(\mathcal{A}).$
\end{Proof}

\begin{Corollary} Let $\mathcal{C}$ be a free $\texttt{R}$-linear codes with generator matrix in
standard form $G,$ and
$G\in\mathcal{M}_{k\times\ell}(\Gamma(\texttt{R})).$ Then
$\texttt{wt}(\pi(\mathcal{C}))=\texttt{wt}(\mathcal{C})$ and
$\pi(G)$ is a generator matrix in standard form for
$\pi(\mathcal{C}).$
\end{Corollary}

\begin{Example} Let $\mathcal{C}$ be the $\mathbb{Z}_{8}$-linear code
 with generator matrix $\left(
\begin{array}{cccccc}
  1 & 1 & 3 & 4 & 0 & 5   \\
  0 & 2 & 2 & 6 & 4 & 0   \\
  0 & 0 & 4 & 0 & 4 & 4
\end{array}
\right).$ The matrix $G$ is in standard form. So the type of
$\mathcal{C}$ is $(1,1,1).$ By Remark\,\ref{Rquot}, a generator
matrix in standard form of
$\mathcal{A}:=\texttt{Annih}_{\mathcal{C}}(2)$ is $\left(
\begin{array}{ccccccc}
  4 & 0 & 0 & 4 & 0 & 4   \\
  0 & 4 & 0 & 4 & 4 & 4   \\
  0 & 0 & 4 & 0 & 4 & 4
\end{array}
\right).$ So by Theorem\,\ref{dis}, we have
$\texttt{wt}(\mathcal{C})=3.$
\end{Example}

\subsection{Galois closure of a linear code over a finite chain ring}

  Let $\mathcal{B}$  be an $\texttt{S}$-linear code of length $\ell.$ Then
$\sigma(\mathcal{B}):=\left\{(\sigma(\textbf{c}_0),\cdots,\sigma(\textbf{c}_{\ell-1}))\;:\;(\textbf{c}_0,\cdots,\textbf{c}_{\ell-1})\in\mathcal{B}\right\}$
is also an $\texttt{S}$-linear code of length $\ell.$  We say that
the $\texttt{S}$-linear code $\mathcal{B}$ is called \index{Galois
invariant}\emph{$\sigma$-invariant} if $\sigma(\mathcal{B}) =
\mathcal{B}.$ The \index{subring subcode}\emph{subring subcode} of
$\mathcal{B}$ to $\texttt{R},$ is an $\texttt{R}$-linear code
$\texttt{Res}_\texttt{R}(\mathcal{B}):=\mathcal{B}\cap
\texttt{R}^{\ell},$ and the \index{trace code}\emph{trace code} of
$\mathcal{B}$ over $\texttt{R},$ is the $\texttt{R}$-linear code
$$
\texttt{Tr}_\texttt{R}^\texttt{S}(\mathcal{B}):=\left\{\left(\texttt{Tr}_\texttt{R}^\texttt{S}(\textbf{c}_0),\cdots,\texttt{Tr}_\texttt{R}^\texttt{S}(\textbf{c}_{\ell-1})\right)\;:\;(\textbf{c}_0,\cdots,\textbf{c}_{\ell-1})\in\mathcal{B}\right\}.
$$
It is clear that
$\texttt{Tr}_\texttt{R}^\texttt{S}(\sigma(\mathcal{B}))=\texttt{Tr}_\texttt{R}^\texttt{S}(\mathcal{B}).$
The \emph{extension code} of an $\texttt{R}$-linear code
$\mathcal{C}$ to $\texttt{S},$ is the $\texttt{S}$-linear code
$\texttt{Ext}_\texttt{S}(\mathcal{C}),$  formed by taking all
combinations of codewords of $\mathcal{C}.$ The following theorem
generalizes Delsarte's celebrated result (see \cite[Ch.7.\S8.
Theorem 11.]{WS77}).

\begin{Theorem} (\cite[Theorem 3]{MNR13}). Let $\mathcal{B}$ be an $\texttt{S}$-linear code then
$\texttt{Tr}_\texttt{R}^\texttt{S}(\mathcal{B}^\perp)
=\texttt{Res}_\texttt{R}(\mathcal{B})^\perp,$ where
$\mathcal{B}^\perp$ is the dual to $\mathcal{B}$ with respect to
the usual scalar product, and
$\texttt{Res}_\texttt{R}(\mathcal{B})^\perp$ is the dual of
$\texttt{Res}_\texttt{R}(\mathcal{B})$ in $\texttt{R}^\ell.$
\end{Theorem}

\begin{Definition}\label{G-inv} Let $\mathcal{B}$ be an $\texttt{S}$-linear
code. The \index{Galois closure}\emph{$\sigma$-closure} of
$\mathcal{B},$ is the smallest $\sigma$-invariant
$\texttt{S}$-linear code $\widetilde{\mathcal{B}},$ containing
$\mathcal{B}.$
\end{Definition}

\begin{Proposition}\label{tra-inv} Let $\mathcal{B}$ be an $\texttt{S}$-linear code.
Then
$\widetilde{\mathcal{B}}=\sum\limits_{i=0}^{m-1}\sigma^i(\mathcal{B})$
and
$\texttt{Tr}_{\texttt{R}}^{\texttt{S}}(\mathcal{B})=\texttt{Tr}_{\texttt{R}}^{\texttt{S}}(\widetilde{\mathcal{B}}).$
\end{Proposition}

\begin{Proof}  We have
 $\mathcal{B}\subseteq\widetilde{\mathcal{B}}$ and $\sigma(\widetilde{\mathcal{B}})=\widetilde{\mathcal{B}},$ by Definition\,\ref{G-inv} of
$\widetilde{\mathcal{B}}.$ So
$\sigma^i(\mathcal{B})\subseteq\widetilde{\mathcal{B}},$ for all
$i\in\{0,1,\cdots,m-1\}.$ Hence
$\sum\limits_{i=0}^{m-1}\sigma^i(\mathcal{B})\subseteq\widetilde{\mathcal{B}}.$
Since
$\sigma\left(\sum\limits_{i=0}^{m-1}\sigma^i(\mathcal{B})\right)=\sum\limits_{i=0}^{m-1}\sigma^i(\mathcal{B})$
and
$\mathcal{B}\subseteq\sum\limits_{i=0}^{m-1}\sigma^i(\mathcal{B}),$
as $\widetilde{\mathcal{B}}$ is the smallest $\texttt{S}$-linear
code containing $\mathcal{B},$ which is $\sigma$-invariant, it
follows
$\widetilde{\mathcal{B}}\subseteq\sum\limits_{i=0}^{m-1}\sigma^i(\mathcal{B}).$
Hence
$\widetilde{\mathcal{B}}=\sum\limits_{i=0}^{m-1}\sigma^i(\mathcal{B}).$
Thanks to \cite[Proposition 1.]{MNR13},
 $\texttt{Tr}_{\texttt{R}}^{\texttt{S}}(\widetilde{\mathcal{B}}) =\texttt{Tr}_{\texttt{R}}^{\texttt{S}}(\mathcal{B}).$
\end{Proof}

The following Theorem summarizes the obtained results  in
\cite{MNR13}.

\begin{Theorem}\label{trace} Let $\mathcal{B}$ be an $\texttt{S}$-linear code and $\sigma$ be a generator of $\texttt{Aut}_{\texttt{R}}(\texttt{S})$. Then the following statements are equivalent:
\begin{enumerate}
     \item $\mathcal{B}$ is  $\sigma$-invariant;
    \item $\texttt{Tr}_{\texttt{R}}^{\texttt{S}}(\mathcal{B})=\texttt{Res}_\texttt{R}(\mathcal{B});$
    \item $\mathcal{B},$ and $\texttt{Res}_\texttt{R}(\mathcal{B})$  have the same type;
    \item
    $\texttt{Res}_\texttt{R}(\mathcal{B})^\perp=\texttt{Res}_\texttt{R}(\mathcal{B}^\perp).$
\end{enumerate}
\end{Theorem}

\begin{Proof} Let $\mathcal{B}$ be an $\texttt{S}$-linear code.
\begin{description}
    \item[$1.\Leftrightarrow 2.$] Thanks to \cite[Theorem 2]{MNR13}.
    \item[$1.\Leftrightarrow 3.$] Since any $\texttt{R}$-basis of $\texttt{Res}_\texttt{R}(\mathcal{B})$ is also an
$\texttt{S}$-basis of
$\texttt{Ext}_\texttt{S}\left(\texttt{Res}_\texttt{R}(\mathcal{B})\right).$
Thanks to \cite[Theorem 1]{MNR13}, we deduce that
$\mathcal{B}=\texttt{Ext}_\texttt{S}\left(\texttt{Tr}_{\texttt{R}}^{\texttt{S}}(\mathcal{B})\right)$
if and only if $\mathcal{B}$ and
$\texttt{Res}_\texttt{R}(\mathcal{B})$ have the same type.
   \item[$2.\Leftrightarrow 4.$] By Delsarte's Theorem, we have $\texttt{Res}_\texttt{R}(\mathcal{B}^\perp)=\texttt{Tr}_{\texttt{R}}^{\texttt{S}}(\mathcal{B})^\perp.$
Thus,
$\texttt{Tr}_{\texttt{R}}^{\texttt{S}}(\mathcal{B})=\texttt{Res}_\texttt{R}(\mathcal{B})$
is equivalent to
$\texttt{Res}_\texttt{R}(\mathcal{B}^\perp)=\texttt{Res}_\texttt{R}(\mathcal{B})^\perp.$
\end{description}
\end{Proof}

\section{Cyclotomic partitions}\label{Sect:4}

Let $\ell$ be a positive integer and $q$ a power of a prime number
with the property $\texttt{gcd}(\ell,q)=1.$

Consider $\mathbb{Z}_\ell,$ the ring of integers modulo $\ell$ and
the underlying set $\Sigma_\ell:=\{0,1,\cdots,\ell-1\}$ of
$\mathbb{Z}_\ell.$ Let $\mathrm{A}$ be a subset of $\Sigma_\ell.$
The set of \emph{multiples} of $u$ in $\mathrm{A}$ is
$u\mathrm{A}:=\{uz\,(\texttt{mod}\,\ell)\;:\;z\in \mathrm{A}\}.$
The \index{Galois closure of a set} $q$-\emph{closure} of a subset
$\mathrm{A}$ of $\Sigma_\ell,$ is
$\complement_q(\mathrm{A}):=\underset{i\in \mathbb{N}}{\cup
}q^i\mathrm{A}.$ We remark that
$\complement_q(\emptyset)=\emptyset.$

\begin{Definition} Let $z\in\Sigma_\ell.$
The \index{cyclotomic coset}\emph{$q$-cyclotomic coset modulo
$\ell,$} containing $z,$ denoted $\complement_q(z),$ is the Galois
closure of $\{z\}.$
\end{Definition}

One denotes by $\Re_\ell(q)$ the set of $q$-closure subsets of
$\Sigma_\ell,$ and by $2^{\Sigma_\ell}$ the set of subsets of
$\Sigma_\ell.$ Obviously, the $q$-cyclotomic cosets modulo $\ell,$
form a partition of $\Sigma_\ell.$ Let $\Sigma_\ell(q)$ be a set
of representatives of each $q$-cyclotomic cosets modulo $\ell.$

\begin{Proposition}\cite[Proposition 5.2]{BGG14}\label{ct-q} We have  $|\Sigma_\ell(q)|
=\sum\limits_{d|\ell}\frac{\phi(\ell)}{\texttt{ord}_\ell(q)},$
where $\phi(.)$ is the Euler function and
$\texttt{ord}_\ell(q):=\texttt{min}\left\{i\in\mathbb{N}\setminus\{0\}\;:\;q^{i}\,\equiv\,1\,(\texttt{mod}\;\ell)\right\}.$
\end{Proposition}

We introduce the binary and unary operations on $\Sigma_\ell.$
These operations are necessary in the following section, for the
construction of cyclic linear codes.

\begin{Definition} Let $z\in\Sigma_\ell$ and $\mathrm{A},\mathrm{B}$ be two
subsets of $\Sigma_\ell.$
\begin{enumerate}
     \item The \index{opposite of an element}\emph{opposite} of $\mathrm{A}$ is $-\mathrm{A}:=\{\ell-z\;:\;z\in \mathrm{A}\}.$
     \item The \index{complementary of a set }\emph{complementary} of $\mathrm{A}$ is $\overline{\mathrm{A}}:=\left\{z\in\Sigma_\ell\;:\;z\not\in \mathrm{A}\right\}.$
     \item The \index{dual of a set}\emph{dual} of $\mathrm{A}$ is $\mathrm{A}^{\diamond}:=\overline{-\mathrm{A}}.$
          \end{enumerate}
\end{Definition}

Let $\mathrm{L}$ be a nonempty set.  We recall that the quintuple
$\langle \mathrm{L};\vee,\wedge;0,1\rangle$ is a \emph{bounded
lattice} if the following identities are satisfied :
\begin{enumerate}
    \item  for all $x,y\in \mathrm{L};$ $x\vee y=y\vee x$ and $x\wedge y=y\wedge x;$
    \item for all $x,y,z\in \mathrm{L};$  $(x\vee y)\vee z=x\vee(y\vee z)$ and $(x\wedge y)\wedge z=x\wedge(y\wedge z);$
    \item  for every $x\in \mathrm{L};$  $x\wedge x=x$ and $x\vee x=x;$
    \item  for every $x\in \mathrm{L};$ $x=x\vee(x\wedge x)$ and $x=x\wedge (x\vee x);$
    \item for every $x\in \mathrm{L},$ $x\wedge 0=0$ and $x\vee 1=1.$
\end{enumerate}
Moreover, a lattice $\langle \mathrm{L};\vee,\wedge\rangle$ is
\emph{distributive} if for all $x,y,z\in \mathrm{L},$ $(x\vee
y)\wedge z=(x\wedge z)\vee(y\wedge z),$ and  a lattice $\langle
\mathrm{L};\vee,\wedge\rangle$ is \emph{modular} if for all
$x,y,z\in \mathrm{L},$ $x\wedge(y\vee(x\vee z))=(x\wedge
z)\vee(y\wedge z).$ A more general and detailed treatment of the
topic can be found in textbooks on Lattices such as \cite{Gra09}.

\begin{Example}
The lattice $\left\langle
2^{\mathrm{E}};\cup,\cap;\emptyset,\mathrm{E}\right\rangle$ is
distributive and bounded, where $2^{\mathrm{E}}$ is the power set
of a set $\mathrm{E}.$  The lattice $\left\langle
\mathcal{L}_\ell(\texttt{R});+,\cap;\{\underline{\textbf{0}}\},\texttt{R}^\ell\right\rangle$
is modular and bounded, where $\mathcal{L}_\ell(\texttt{R})$ is
the set of all $\texttt{R}$-linear codes of length $\ell.$
\end{Example}

The relationships among these operations, are given in the
following:

\begin{Proposition} The lattice $\left\langle\,\Re_\ell(q); \cup,\cap;\emptyset,\Sigma_\ell(q)\,\right\rangle$ is bounded and distributive.
Moreover, the map $$\begin{array}{cccc}
         \complement_q: &  2^{\Sigma_\ell} & \rightarrow & \Re_\ell(q) \\
              & \mathrm{A} & \mapsto & \underset{i\in \mathbb{N}}{\cup }q^i\mathrm{A}
          \end{array}$$
is a lattices epimorphism with
$\complement_q(-\mathrm{A})=-\complement_q(\mathrm{A}),$
$\complement_q(\overline{\mathrm{A}})=\overline{\complement_q(\mathrm{A})},$
and $\mathrm{A}\subseteq\mathrm{B}$ implies
$\complement_q(\mathrm{A})\subseteq\complement_q(\mathrm{B}).$
\end{Proposition}

\begin{Definition}\label{D-cp} The $(s+1)$-tuple
$(\mathrm{A}_0,\mathrm{A}_1,\cdots,\mathrm{A}_s)$ is an
$(q,s)$-\index{cyclotomic partition}\emph{cyclotomic partition}
modulo $\ell,$ if there exists a unique map
$\lambda:\Sigma_\ell(q)\rightarrow\{0,1,\cdots, s\},$ such that
$\mathrm{A}_t=\complement_q\left(\lambda^{-1}(\{t\})\right),$ for
every $0\leq t\leq s.$
\end{Definition}

We denoted by $\Re_\ell(q,s)$ the set of $(q,s)$-cyclotomic
partitions modulo $\ell.$ It is easy to see that
$$\Re_\ell(q,s):=\left\{(\mathrm{A}_0,\mathrm{A}_1,\cdots,\mathrm{A}_s) \;\;:\; \left(\exists
\lambda\in\{0,1,\cdots,s\}^{\Sigma_\ell(q)}\right)\left(\mathrm{A}_t=\lambda^{-1}(\{t\})\right)\right\}$$
 By Definition\,\ref{D-cp}, we see that
$|\Re_\ell(q,s)| =(s+1)^{|\Sigma_\ell(q)|}.$

\begin{Example} We take $\;\ell=20, q=3,$ and $s=2.$ The $q$-cyclotomic cosets modulo
$\ell,$ are: $$ \complement_q(\{0\})=\{0\},
\complement_q(\{5\})=\{5,15\}, \complement_q(\{10\})=\{10\},$$ and
$$\begin{array}{ll}
 \complement_q(\{1\})=\{1,3,9,7\} ; & \complement_q(\{2\})=\{2,6,18,14\}; \\
 \complement_q(\{4\})=\{4,12,16,8\}; &  \complement_q(\{11\})=\{11,13,19,17\}.
\end{array}$$
 So $\Sigma_\ell(q)=\{0,1,2,4,5,10,11\}.$ We remark
that $\complement_q(\{-z\})=\complement_q(\{z\}),$ for every
$z\in\{0,2,4,5,10\}.$ We set $\;\mathrm{I}:=\{0;1;2;\cdots;10\}.$
We have
$\mathrm{A}:=\complement_q(\mathrm{I})=\complement_q(\{0,1,2,4,5,10\}),$
$-\mathrm{A}=\complement_q(\{2,4,5,10,11\}),$ and
$\mathrm{A}^\diamond:=\complement_q(\{1\}).$
\end{Example}

We remark that the maps
$$\begin{array}{cccc}
          \psi : &  \Re_\ell(q)  & \hookrightarrow & \Re_\ell(q,s)  \\
              & \mathrm{A} & \mapsto &
              (\mathrm{A},\emptyset,\cdots,\emptyset,\overline{\mathrm{A}})
          \end{array}~~~~\text{  and  }~~~~\begin{array}{cccc}
          \varphi : &  \Re_\ell(q,s)   & \twoheadrightarrow & \Re_\ell(q) \\
              & (\mathrm{A}_0,\cdots,\mathrm{A}_s) & \mapsto & \mathrm{A}_0
          \end{array}$$
satisfy $\varphi\psi(\mathrm{A})=\mathrm{A},$ and
$\psi\varphi(\mathrm{A}_0,\mathrm{A}_1,\cdots,\mathrm{A}_s)=(\mathrm{A}_0,\emptyset,\cdots,\emptyset,\overline{\mathrm{A}_0})$
for every $\mathrm{A}\in\Re_\ell(q),$ for every
$(\mathrm{A}_0,\mathrm{A}_1,\cdots,\mathrm{A}_s)\in\Re_\ell(q,s).$

\section{Free Cyclic linear codes over finite chain rings}\label{Sect:5}

Let $\texttt{R}$ be a finite chain ring of invariants $(q,s)$ and
$\ell$ be a positive integer such that $\texttt{gcd}(q,\ell)=1.$
Then there exists a positive integer $m$ such that  $q^{m}\equiv
1(\texttt{mod}\,\ell)$ and $q^{m-1}\not\equiv
1(\texttt{mod}\,\ell).$  In this section, we give the trace
representation of cyclic $\texttt{R}$-linear codes of length
$\ell.$

\subsection{Cyclic polynomial codes over finite chain rings}

Let $\texttt{S}$ be the Galois extension of $\texttt{R}$ of degree
$m$ and $\xi$ be a generator of
$\Gamma(\texttt{S})\backslash\{0\}.$  Let
$\mathrm{A}:=\{a_1,a_2,\cdots,a_k\}$ be a subset of $\Sigma_\ell.$
One denotes by $\textbf{P}(\texttt{S}\,;\,\mathrm{A}),$ the free
$\texttt{S}$-submodule of $\texttt{S}[X]$ with free
$\texttt{S}$-basis $\{X^a\,:\,a\in\mathrm{A}\}.$ Since $m$ is the
smallest positive integer with $q^m\equiv 1
\,(\texttt{mod}\,\ell),$ we can write
$\eta:=\xi^{\frac{q^m-1}{\ell}}$ and the multiplicative order of
$\eta$ is  $\ell.$ The evaluation
$$\begin{array}{cccc}
  \textbf{ev}_\eta: & \textbf{P}(\texttt{S}\,;\,\mathrm{A}) & \rightarrow & \texttt{S}^\ell \\
   & f & \mapsto &
   (f(1),f(\eta),\cdots,f(\eta^{\ell-1})),
\end{array}$$ is an $\texttt{S}$-modules monomorphism. We see that if $\mathrm{A}:=\{0,1,\cdots,k-1\},$ then for any  $\ell^{\texttt{th}}$-primitive root of unity
$\eta$ in $\Gamma(\texttt{S}),$ the $\texttt{S}$-linear code
$\textbf{ev}_\eta(\textbf{P}(\texttt{S}\,;\,\mathrm{A}))$ is a
primitive Reed-Solomon code. For this reason, we define cyclic
polynomial codes which is a family of codes over large finite
chain rings as follows.

\begin{Definition} Let  $\mathrm{A}$ be a subset of $\Sigma_\ell,$ and $\texttt{S}$ be a finite chain ring such that $|\Gamma(\texttt{S})|\geq\ell.$
Let $\eta\in\Gamma(\texttt{S})$ with the multiplicative order of
$\eta$ is $\ell.$ The \index{cyclic polynomial code}\emph{cyclic
polynomial code} over $\texttt{S},$ with defining set
$\mathrm{A},$ denoted
$\textbf{L}_\eta(\texttt{S}\,;\,\mathrm{A}),$ is the free
$\texttt{S}$-module
$\textbf{ev}_\eta(\textbf{P}(\texttt{S}\,;\,\mathrm{A})).$
\end{Definition}

For every subset set $\mathrm{A}$ of $\Sigma_\ell,$ and for every
positive integer $u$ such that  $\texttt{gcd}(u,\ell)=1,$ we have
$\textbf{L}_{\eta^u}(\texttt{S}\,;\,\mathrm{A})=\textbf{L}_\eta(\texttt{S}\,;\,u\mathrm{A})$
and \begin{align}\label{won} W_{\mathrm{A}}:=\left(%
\begin{array}{cccc}
  1 & \eta^{a_1} & \cdots & \eta^{(\ell-1)a_1} \\
  \vdots & \vdots &   & \vdots \\
  1 & \eta^{a_k} & \cdots & \eta^{(\ell-1)a_k}
\end{array}%
\right)\end{align} is a generator matrix for
$\textbf{L}_\eta(\texttt{S}\,;\,\mathrm{A}).$ We see that
$\textbf{L}_\eta(\texttt{S}\,;\,\emptyset)=\{\underline{\textbf{0}}\},$
$\textbf{L}_\eta(\texttt{S}\,;\,\{0\})=\textbf{1}$ and
$\textbf{L}_\eta(\texttt{S}\,;\,\Sigma_\ell)=\texttt{S}^\ell.$

\begin{Proposition}\label{cyclic} Let $\mathrm{A}$ be a subset of $\Sigma_\ell.$ Then $\textbf{L}_\eta(\texttt{S}\,;\,\mathrm{A})$ is
cyclic. \end{Proposition}

\begin{Proof} Consider the codeword $\textbf{c}_a =
\left(1,\eta^a, \cdots,\eta^{a(\ell-1)}\right).$ Then the shift of
$\textbf{c}_a$ is $\eta^{-a}\textbf{c}_a.$ Since
$\textbf{L}_\eta(\texttt{S}\,;\,\mathrm{A})$ is
$\texttt{S}$-linear, we have
$\eta^{-a}\textbf{c}_a\in\textbf{L}_\eta(\texttt{S}\,;\,\mathrm{A}).$
Hence $\textbf{L}_\eta(\texttt{S}\,;\,\mathrm{A})$ is cyclic.
\end{Proof}

\begin{Proposition}\label{L-opera} Let $\mathrm{A}, \mathrm{B}$ be two subsets of
$\Sigma_\ell.$   The following assertions holds:
\begin{enumerate}
    \item  $\textbf{L}_\eta(\texttt{S}\,;\,\mathrm{A})^\perp=\textbf{L}_\eta(\texttt{S}\,;\,\mathrm{A}^{\diamond});$
    \item  $\textbf{L}_\eta(\texttt{S}\,;\,\mathrm{A}\cup\mathrm{B})=\textbf{L}_\eta(\texttt{S}\,;\,\mathrm{A})+\textbf{L}_\eta(\texttt{S}\,;\,\mathrm{B})$ and $\textbf{L}_\eta(\texttt{S}\,;\,\mathrm{A}\cap\mathrm{B})=\textbf{L}_\eta(\texttt{S}\,;\,\mathrm{A})\cap\textbf{L}_\eta(\texttt{S}\,;\,\mathrm{B}).$
\end{enumerate}
\end{Proposition}

\begin{Proof} Let $\mathrm{\mathrm{A}}, \mathrm{B}$ be  subsets of
$\Sigma_\ell.$
\begin{enumerate}
    \item An $\texttt{S}$-basis of
$\textbf{L}_\eta(\texttt{S}\,;\,\mathrm{A}^{\diamond})$ is
$\{\textbf{c}_a\;:\;-a\in \overline{\mathrm{A}}\}$ where
$\textbf{c}_a:=(1,\eta^{-a},\cdots,\eta^{-a(\ell-1)})\in\textbf{L}_\eta(\texttt{S}\,;\,\mathrm{A}^{\diamond}).$
Then for all $b\in \mathrm{A},$
$\textbf{c}_b:=(1,\eta^{b},\cdots,\eta^{b(\ell-1)})\in\textbf{L}_\eta(\texttt{S}\,;\,\mathrm{A}),$
we have
$\textbf{c}_b\textbf{c}_a^{\texttt{T}}=\sum\limits_{j=0}^{\ell-1}\eta^{(b-a)j},$
where $\textbf{c}_a^{\texttt{T}}$ is the transpose of
$\textbf{c}_a.$ It is easy to check that
$\sum\limits_{j=0}^{\ell-1}\eta^{ij}=0,$ when $i\not\equiv\,
0(\texttt{mod}\,\ell).$ Since
 $0<b-a<\ell,$ we have $\textbf{c}_b\textbf{c}_a^{\texttt{T}}=0.$ So
$\textbf{L}_\eta(\texttt{S}\,;\,\mathrm{A}^{\diamond})\subseteq\textbf{L}_\eta(\texttt{S}\,;\,\mathrm{A})^\perp.$
Comparison of cardinality yields
$\textbf{L}_\eta(\texttt{S}\,;\,\mathrm{A})^\perp=\textbf{L}_\eta(\texttt{S}\,;\,\mathrm{A}^{\diamond}).$
    \item We have
$\textbf{L}_\eta(\texttt{S}\,;\,\mathrm{A})\subseteq\textbf{L}_\eta(\texttt{S}\,;\,\mathrm{A}\cup
\mathrm{B}),$
$\textbf{L}_\eta(\texttt{S}\,;\,\mathrm{B})\subseteq\textbf{L}_\eta(\texttt{S}\,;\,\mathrm{A}\cup
\mathrm{B}).$ Therefore
$$\textbf{L}_\eta(\texttt{S}\,;\,\mathrm{A})+\textbf{L}_\eta(\texttt{S}\,;\,\mathrm{B})\subseteq\textbf{L}_\eta(\texttt{S}\,;\,\mathrm{A}\cup
\mathrm{B})\text{ and
}\textbf{L}_\eta(\texttt{S}\,;\,\mathrm{A})\cap\textbf{L}_\eta(\texttt{S}\,;\,\mathrm{B})\subseteq\textbf{L}_\eta(\texttt{S}\,;\,\mathrm{A}\cup
\mathrm{B}).$$
 Since an $\texttt{S}$-basis of
$\textbf{L}_\eta(\texttt{S}\,;\,\mathrm{A})+\textbf{L}_\eta(\texttt{S}\,;\,\mathrm{B}),$
is $\{\textbf{c}_a\;:\;a\in \mathrm{A}\cup
(\mathrm{B}\setminus\mathrm{A})\}$ and an $\texttt{S}$-basis of
$\textbf{L}_\eta(\texttt{S}\,;\,\mathrm{A})\cap\textbf{L}_\eta(\texttt{S}\,;\,\mathrm{B}),$
is $\{\textbf{c}_a\;:\;a\in \mathrm{A}\cap \mathrm{B}\}.$ We have
the equalities.
   \end{enumerate}
\end{Proof}

We set  $\textbf{L}_\ell(\texttt{S})$ the set of all cyclic
polynomial codes of length $\ell,$ over $\texttt{S}.$ Then the
quintuple
$\left\langle\,\textbf{L}_\ell(\texttt{S});+,\,\cap;\,\{\underline{\textbf{0}}\},\,\texttt{S}^\ell\right\rangle,$
is a lattice and the map
$$\begin{array}{cccc}
         \textbf{L}_\eta(\texttt{S}\,;\,-): & 2^{\Sigma_\ell} &\rightarrow & \textbf{L}_\ell(\texttt{S}) \\
              & \mathrm{A} & \mapsto & \textbf{L}_\eta(\texttt{S}\,;\,\mathrm{A})
          \end{array}$$
is a bijective lattices homomorphism. The following result extends
\cite[Theorem 5]{Bie02} to finite chain rings.

\begin{Definition} A subset $\mathrm{I}$ of $\Sigma_\ell$ is an interval of
length $\delta$ if there exists
$(a,u)\in\Sigma_\ell\times\Sigma_\ell$ such that
$\texttt{gcd}(u,\ell)=1$ and
$\mathrm{I}:=\biggl\{ua,\,u(a+1),\,\cdots,u(a+\delta-1)\,\biggr\}.$
\end{Definition}

\begin{Theorem}\label{bch} If $\;\mathrm{A}^{\diamond}\;$ contains an interval of length
$\delta$ then
$\texttt{wt}\left(\textbf{L}_\eta(\texttt{S}\,;\,\mathrm{A})\right)\geq
\delta+1.$
\end{Theorem}

\begin{Proof} Let
$\mathrm{I}=\biggl\{ua,u(a+1),\cdots,u(a+\delta-1)\biggr\}$ be an
interval containing in $\;\mathrm{A}^{\diamond}\;,$ for some
integer $u\in\Sigma_\ell$ such that $(u,\ell)=1.$ Then
$\zeta:=\eta^u$ is also a primitive root of $X^\ell-1.$  Suppose
that $\textbf{c}$ is a nonzero codeword of
$\textbf{L}_\eta(\texttt{S}\,;\,u^{-1}A)$ with
$\texttt{wt}(\textbf{c})=\texttt{wt}(\textbf{L}_\eta(\texttt{S}\,;\,u^{-1}\texttt{A}))$
and
$\texttt{wt}(\textbf{L}_\eta(\texttt{S}\,;\,u^{-1}\texttt{A}))=\texttt{wt}(\textbf{L}_\eta(\texttt{S}\,;\,\mathrm{A})).$
Since $W_{\mathrm{B}}$ is a parity-check matrix of
$\textbf{L}_\eta(\texttt{S}\,;\,A)$ and
$\mathrm{B}:=\mathrm{A}^{\diamond},$ it follows that
$W_{\mathrm{B}}\textbf{c}^{\texttt{T}}=\underline{\textbf{0}}.$
Assume that $\texttt{Supp}(\textbf{c}):=\{j\;:\;c_j\neq
0\}\subseteq\{j_1,j_2,\cdots,j_\delta\}.$ Consider
$\textbf{m}:=(c_{j_1},c_{j_2},\cdots,c_{j_\delta})$ where
$c_{j_1},c_{j_2},\cdots,c_{j_\delta}$ are the first $\delta$
entries of
$\textbf{c}=(\cdots,0,c_{j_i},0,\cdots,0,c_{j_{i+1}},0,\cdots).$
Thus the equality
$W_{\mathrm{B}}\textbf{c}^{\texttt{T}}=\underline{\textbf{0}}$
becomes
$W\textbf{m}^{\texttt{T}}=\underline{\textbf{0}},$ where $$W_{\underline{\delta}}:=\left(%
\begin{array}{cccc}
  \zeta^{j_1a} & \cdots & \zeta^{j_\delta a}   \\
  \vdots &   & \vdots   \\
  \zeta^{j_1(a+\delta-1)} & \cdots & \zeta^{j_\delta(a+\delta-1)}
\end{array}%
\right).$$  We have
$\texttt{det}(W_{\underline{\delta}})=-\zeta^{\delta\left(\sum\limits_{t=1}^\delta
j_t\right)}\underset{1\leq u<w\leq
\delta}{\prod}\left(\zeta^{j_v}-\zeta^{j_w}\right)$ is invertible
since $\zeta\in\Gamma(\texttt{S})\setminus\{0\}.$ Therefore
$\textbf{m}=\underline{\textbf{0}}$ which is a contradiction
because $\textbf{c}\neq\underline{\textbf{0}}.$ Hence
$\texttt{wt}(\textbf{L}_\eta(\texttt{S}\,;\,\mathrm{A}))\geq
\delta+1.$
\end{Proof}

\begin{Proposition} Let $\texttt{S}$ be a finite chain ring of invariants $(2^n,s)$ and $\;\ell:=2^{sn}-1,$
$\mathrm{A}:=\{1,2,\cdots,d-1\}$ where $d:=2^{sn-1}.$ Then
$\textbf{L}_\eta(\texttt{S}\,;\,\mathrm{A}^{\diamond})$ is MDS and
self-orthogonal.
\end{Proposition}

\begin{Proof} We have $\mathrm{A}$ is an interval of length $d-1$ and
$\mathrm{A}^{\diamond}:=\mathrm{A}\cup\{0\}.$ So We have
$\textbf{L}_\eta(\texttt{S}\,;\,\mathrm{A})^\perp=\textbf{L}_\eta(\texttt{S}\,;\,\mathrm{A}^{\diamond})=\textbf{L}_\eta(\texttt{S}\,;\,\mathrm{A})\oplus\textbf{\textbf{1}}.$
Thus $\textbf{L}_\eta(\texttt{S}\,;\,\mathrm{A}^{\diamond})$ is
self-orthogonal and $\mathrm{A}^{\diamond}$ is also an interval of
length $d.$ Thanks to BCH-bound of Theorem\,\ref{bch}, we see that
$\textbf{L}_\eta(\texttt{S}\,;\,\mathrm{A}^{\diamond})$ is MDS.
\end{Proof}

\subsection{Trace representation of free cyclic linear codes}

We consider the trace-evaluation
 $\texttt{Tr}_\texttt{R}^\texttt{S}\circ\textbf{ev}_\eta:\textbf{P}_\eta(\texttt{S};\mathrm{A})\rightarrow \texttt{R}^\ell,$ defined
 by:
 $$\texttt{Tr}_\texttt{R}^\texttt{S}\circ\textbf{ev}_\eta(\textbf{b}X^a):=\left(\texttt{Tr}_\texttt{R}^\texttt{S}(\textbf{b}),\texttt{Tr}_\texttt{R}^\texttt{S}(\textbf{b}\eta^a),\cdots,\texttt{Tr}_\texttt{R}^\texttt{S}(\textbf{b}\eta^{a(\ell-1)})\right),$$
 for every $a\in  \mathrm{A}$ and for every $\textbf{b}\in
\texttt{R}.$ In the sequel, we write:
 $\mathrm{\textbf{C}}_\eta(\texttt{R}\,;\,\mathrm{A}):=\texttt{Tr}_\texttt{R}^\texttt{S}\left(\textbf{L}_\eta(\texttt{S}\,;\,\mathrm{A})\right),$
 and  $\mathrm{\textbf{C}}_\eta(\texttt{R}\,;\,\mathrm{A})$ is a free cyclic $\texttt{R}$-linear code of length $\ell.$
 Let
$\textbf{C}_\ell(\texttt{R}):=\left\{\mathrm{\textbf{C}}_\eta(\texttt{R}\,;\,\mathrm{A})\;:\;\mathrm{A}\subseteq\Sigma_\ell\right\}$
be the set of all free cyclic $\texttt{R}$-linear codes of length
$\ell.$

 \begin{Proposition}\label{c(a)}
 Let $\mathrm{\mathrm{A}}, \mathrm{B}$ be two subsets of $\Sigma_\ell.$  Then
 \begin{enumerate}
     \item  $\textbf{L}_\eta\left(\texttt{S}\,;\,\complement_q(\mathrm{A})\right)$ is the $\sigma$-closure of $\textbf{L}_\eta(\texttt{S}\,;\,\mathrm{A})$ and $\mathrm{\textbf{C}}_\eta(\texttt{R}\,;\,\mathrm{A})=\mathrm{\textbf{C}}_\eta\left(\texttt{R}\,;\,\complement_q(\mathrm{A})\right);$
     \item $\texttt{rank}_{\texttt{S}}(\textbf{L}_\eta(\texttt{S}\,;\,\complement_q(\mathrm{A})))=|\complement_q(\mathrm{A})|;$
     \item $\mathrm{\textbf{C}}_\eta(\texttt{R}\,;\,\mathrm{A})^{\perp}=\mathrm{\textbf{C}}_\eta(\texttt{R}\,;\,\mathrm{A}^{\diamond});$
     \item $\mathrm{\textbf{C}}_\eta\left(\texttt{R}\,;\,\mathrm{A}\cup\mathrm{B}\right)=\mathrm{\textbf{C}}_\eta\left(\texttt{R}\,;\,\mathrm{A}\right)+\mathrm{\textbf{C}}_\eta\left(\texttt{R}\,;\,\mathrm{B}\right)$ and $\mathrm{\textbf{C}}_\eta\left(\texttt{R}\,;\,\mathrm{A}\cap\mathrm{B}\right)=\mathrm{\textbf{C}}_\eta\left(\texttt{R}\,;\,\mathrm{A}\right)\cap\mathrm{\textbf{C}}_\eta\left(\texttt{R}\,;\,\mathrm{B}\right).$
    \end{enumerate}
  \end{Proposition}

\begin{Proof} Let $\mathrm{A}, \mathrm{B }$ be two subsets of
$\Sigma_\ell.$
\begin{enumerate}
    \item On the one hand, it is clear that  $\sigma(\textbf{L}_\eta(\texttt{S}\,;\,\mathrm{A}))=\textbf{L}_\eta(\texttt{S}\,;\,q\mathrm{A}).$
    So by Proposition\,\ref{tra-inv}, we have
    $$\widetilde{\textbf{L}_\eta(\texttt{S}\,;\,\mathrm{A})}=\sum\limits_{i=0}^{m-1}\textbf{L}_\eta(\texttt{S}\,;\,q^i\mathrm{A})=\textbf{L}_\eta\left(\texttt{S}\,;\,\overset{m-1}{\underset{i=0}{\bigcup}}q^i\mathrm{A}\right).$$
    Since $\complement_q(\mathrm{A})=\overset{m-1}{\underset{i=0}{\bigcup}}q^i\mathrm{A},$ we obtain
    $\widetilde{\textbf{L}_\eta(\texttt{S}\,;\,\mathrm{A})}=\textbf{L}_\eta(\texttt{S}\,;\,\complement_q(\mathrm{A}))$ and the other hand,
    from Proposition\,\ref{tra-inv},
$\mathrm{\textbf{C}}_\eta(\texttt{R}\,;\,\mathrm{A})=\texttt{Tr}(\textbf{L}_\eta(\texttt{S}\,;\,\mathrm{A}))=\texttt{Tr}(\textbf{L}_\eta(\texttt{S}\,;\,\complement_q(\mathrm{A})))=\mathrm{\textbf{C}}_\eta(\texttt{R}\,;\,\complement_q(\mathrm{A})).$
    \item Theorem\,\ref{trace}(3) yields $\mathrm{\textbf{C}}_\eta(\texttt{R}\,;\,\mathrm{A})=\texttt{Tr}(\textbf{L}_\eta(\texttt{S}\,;\,\complement_q(\mathrm{A})))=\texttt{Res}_\texttt{R}(\textbf{L}_\eta(\texttt{S}\,;\,\complement_q(\mathrm{A}))).$
    So $$\texttt{rank}_{\texttt{R}}(\mathrm{\textbf{C}}_\eta(\texttt{R}\,;\,\mathrm{A}))=\texttt{rank}_{\texttt{S}}(\textbf{L}_\eta(\texttt{S}\,;\,\complement_q(\mathrm{A})))=|\complement_q(\mathrm{A})|.$$
     \item We have
\begin{eqnarray*}
  \mathrm{\textbf{C}}_\eta(\texttt{R}\,;\,\mathrm{A})^{\perp} &=& \texttt{Res}_\texttt{R}\left(\textbf{L}_\eta(\texttt{S}\,;\, \complement_q(\mathrm{A}))^{\perp}\right), \text{by Theorem\,\ref{trace}} ; \\
   &=& \texttt{Res}_\texttt{R}\left(\textbf{L}_\eta(\texttt{S}\,;\,\complement_q(\mathrm{A})^\diamond)\right), \text{by Proposition\,\ref{L-opera}} ; \\
    &=& \mathrm{\textbf{C}}_\eta\left(\texttt{R}\,;\, \mathrm{A}^\diamond\right).
\end{eqnarray*}
     \item   Thanks to Proposition\,\ref{L-opera}, $\mathrm{\textbf{C}}_\eta(\texttt{R}\,;\,\mathrm{A}\cup \mathrm{B})=\texttt{Tr}(\textbf{L}_\eta(\texttt{S}\,;\,\mathrm{A}\cup \mathrm{B}))=\texttt{Tr}(\textbf{L}_\eta(\texttt{S}\,;\,\mathrm{A}))+\texttt{Tr}(\textbf{L}_\eta(\texttt{S}\,;\,\mathrm{B})).$
    Since $\complement_q(\mathrm{A})\cap \complement_q(\mathrm{B})=\emptyset,$ we have $\texttt{Tr}(\textbf{L}_\eta(\texttt{S}\,;\,\mathrm{A}))\cap\texttt{Tr}(\textbf{L}_\eta(\texttt{S}\,;\,\mathrm{B}))=\{\underline{\textbf{0}}\}.$
\end{enumerate}
\end{Proof}

 \begin{Theorem}\label{L-fcy}  Let $\ell,q$ be positive integers such that $q$ is a prime power and $\texttt{gcd}(q,\ell)=1.$
 Then the lattices
$\left\langle\,\textbf{C}_\ell(\texttt{R});+,\cap;
\{\underline{\textbf{0}}\} ,\texttt{R}^\ell\,\right\rangle$ and
$\left\langle\,\Re_\ell(q);
\cup,\cap;\emptyset,\Sigma_\ell(q)\,\right\rangle$ are isomorphic.
\end{Theorem}

 \begin{Proof} It is obvious that prove that the map
 $\mathrm{\textbf{C}}_\eta(\texttt{R}\,;\-):\Re_\ell(q)\rightarrow\textbf{C}_\ell(\texttt{R}),$
 is a bijective.
 From Proposition\,\ref{c(a)}, this map is a lattice isomorphism.
\end{Proof}

 \begin{Corollary}  Let $\ell,q$ be positive integers such that $q$ is a prime power and $\texttt{gcd}(q,\ell)=1.$  Then the
 lattices
 $\left\langle\texttt{Cy}(\mathbb{F}_q,\ell);\;+,\cap;\{\underline{\textbf{0}}\},\mathbb{F}_q^\ell\right\rangle$
 and $\left\langle\,\textbf{C}_\ell(\texttt{R});+,\cap; \{\underline{\textbf{0}}\}
,\texttt{R}^\ell\,\right\rangle$ are isomorphic.
\end{Corollary}

 \begin{Proof} For all finite chain rings $\texttt{R}_1$ and $\texttt{R}_2$ of invariants $(q,s_1)$ and $(q,s_2)$ respectively. By Theorem \ref{L-fcy}, lattices
$\left\langle\,\textbf{C}_\ell(\texttt{R}_1);+,\cap;
\{\underline{\textbf{0}}\} ,\texttt{R}_1^\ell\,\right\rangle$ and
$\left\langle\,\textbf{C}_\ell(\texttt{R}_2);+,\cap;
\{\underline{\textbf{0}}\} ,\texttt{R}_2^\ell\,\right\rangle$ are
isomorphic. Since
$\textbf{C}_\ell(\mathbb{F}_q)=\texttt{Cy}(\mathbb{F}_q,\ell),$ we
have the result.
\end{Proof}

 \begin{Lemma}\label{irr1} Let $\texttt{R}$ be a finite chain ring of invariants $(q,s)$ and $\texttt{S}$ be the Galois extension of $\texttt{R}$
 of degree $m.$ Let $z\in\Sigma_\ell.$
 Set $\texttt{S}=\texttt{R}[\xi],$ $m_z:=|\complement_q(z)|,$ $\eta:=\xi^{\frac{q^m-1}{\ell}}.$ and  $\zeta:=\eta^{-z}.$  Then the map
$$\begin{array}{cccc}
  \psi_{z}: & \texttt{R}[\xi^{z}] & \longrightarrow & \mathrm{\textbf{C}}_\eta\left(\texttt{R}\,;\,\{z\}\right) \\
      & \textbf{a} & \longmapsto & \texttt{Tr}_\texttt{R}^\texttt{S}\left(\emph{ev}_\eta(\textbf{a}X^{z})\right) \\
  \end{array}
$$ is an $\texttt{R}$-module isomorphism. Further $\psi_{z}\circ t_{\zeta}=\tau_1\circ\psi_{z},$
where $\tau_1$ is the cyclic shift and
$t_\zeta(\textbf{a})=\textbf{a}\zeta,$ for all
$\textbf{a}\in\texttt{R}[\eta].$
\end{Lemma}

\begin{Proof} It is
clear that  $\textbf{a}\in\texttt{Ker}(\psi_{z})$ if and only if
$\textbf{a}\in\texttt{R}[\xi^{z}]^{\perp_{\texttt{Tr}}}\cap\texttt{R}[\xi^{z}],$
where duality $\perp_{\texttt{Tr}}$ is with respect to trace form.
As the trace bilinear form is nondegenerate, we have
$\texttt{S}=\texttt{R}[\xi^{z}]^{\perp_{\texttt{Tr}}}\oplus\texttt{R}[\xi^{z}]$
and $\texttt{Ker}(\psi_{z})=\{0_\texttt{R}\}.$ Hence $\psi_{z}$ is
an $\texttt{R}$-module monomorphism. We remark that,
$\mathrm{\textbf{C}}_\eta(\texttt{R}\,;\,\{z\})$ is cyclic, if and
only if $\psi_{z}\circ t_{\zeta}=\tau_1\circ\psi_{z}.$ Finally, we
have $\texttt{S}=\texttt{R}[\xi],$ and by Proposition\,\ref{Gsr},
the ring $\texttt{R}[\xi^{z}]$ is the Galois extension of
$\texttt{R}$ of degree $m_z.$ Hence, $\psi_{z}$ is an
$\texttt{R}$-module isomorphism.
\end{Proof}

\begin{Definition} A non trivial cyclic $\texttt{R}$-linear code
$\mathcal{C}$ is said to be \index{irreducible cyclic linear
code}\emph{irreducible}, if for all $\texttt{R}$-linear cyclic
subcodes $\mathcal{C}_1$ and $\mathcal{C}_2$ of $\mathcal{C},$ the
implication holds: $\mathcal{C}=\mathcal{C}_1\oplus\mathcal{C}_2,$
then $\mathcal{C}_1=\{\underline{\textbf{0}}\}$ or $
\mathcal{C}_2=\{\underline{\textbf{0}}\}.$
\end{Definition}

\begin{Proposition}\label{irr(a)}
The irreducible cyclic $\texttt{R}$-linear codes are precisely
$\theta^t\mathrm{\textbf{C}}_\eta(\texttt{R}\,;\,\{z\})$s, where
$t\in\{0,1,\cdots,s-1\}$ and  $z\in\Sigma_\ell(q).$
\end{Proposition}

\begin{Proof} By Lemma \ref{irr1}, the free cyclic $\texttt{R}$-linear codes of length $\ell$ are $\mathrm{\textbf{C}}_\eta(\texttt{R}\,;\,\{z\}))$
where $z\in\{0,1,\cdots,\ell-1\}$ and all the $\texttt{R}$-linear
cyclic subcodes of each cyclic $\texttt{R}$-linear code, are
irreducible. Let $\mathcal{C}$ be an irreducible cyclic
$\texttt{R}$-linear code. Then
$\mathcal{A}:=\texttt{Annih}_\mathcal{C}(\theta)$ is also an
irreducible cyclic $\texttt{R}$-linear code and the
$\texttt{R}$-linear code $\texttt{Quot}_{s-1}(\mathcal{A})$ is
cyclic and free. Let $\texttt{Quot}_{s-1}(\mathcal{A})$ be the
free cyclic $\texttt{R}$-linear code such that
$\mathcal{A}\subset\texttt{Quot}_{s-1}(\mathcal{A})$ and
$\texttt{rank}_\texttt{R}(\mathcal{A})=\texttt{rank}_\texttt{R}(\texttt{Quot}_{s-1}(\mathcal{A})).$
Assume that $|\mathrm{A}|>1.$ Then
$\mathrm{\textbf{C}}_\eta(\texttt{R}\,;\,\mathrm{A})=\mathrm{\textbf{C}}_\eta(\texttt{R}\,;\,\mathrm{A}_1)\oplus\mathrm{\textbf{C}}_\eta(\texttt{R}\,;\,\mathrm{A}_2)$
where $\mathrm{A}_1\cap \mathrm{A}_2=\emptyset,$
$\mathrm{A}_1\neq\emptyset$ and $\mathrm{A}_2\neq\emptyset.$ We
have
$\mathcal{C}\cap\mathrm{\textbf{C}}_\eta(\texttt{R}\,;\,\mathrm{A}_1)\neq\{\underline{\textbf{0}}\}$
and
$\mathcal{C}\cap\mathrm{\textbf{C}}_\eta(\texttt{R}\,;\,\mathrm{A}_2)\neq\{\underline{\textbf{0}}\}.$
Therefore
$\mathcal{C}=(\mathcal{C}\cap\mathrm{\textbf{C}}_\eta(\texttt{R}\,;\,\mathrm{A}_1))\oplus(\mathcal{C}\cap\mathrm{\textbf{C}}_\eta(\texttt{R}\,;\,\mathrm{A}_2)).$
It is impossible, because $\mathcal{C}$ be an irreducible. So
$|\mathrm{A}|=1.$ Now,
$\mathcal{C}\subseteq\mathrm{\textbf{C}}_\eta(\texttt{R}\,;\,\{z\}),$
it follows that
$\mathcal{C}=\theta^t\mathrm{\textbf{C}}_\eta(\texttt{R}\,;\,\{z\}),$
for some $t\in\{0,1,\cdots,s-1\}.$
\end{Proof}

\section{Sum and intersection of cyclic linear codes}\label{Sect:6}

Consider the map
\begin{align}\label{map0}
\begin{array}{cccc}
 \mathrm{\textbf{C}}_{\ell,\texttt{R}} :  & ~~\Re_\ell(q,s)~~& \rightarrow & \texttt{Cy}(\texttt{R},\ell) \\
   &  (\mathrm{A}_0,\mathrm{A}_1,\cdots,\mathrm{A}_s) & \mapsto &
   \overset{s-1}{\underset{t=0}{\bigoplus}}\theta^{t}\mathrm{\textbf{C}}_\eta(\texttt{R}\,;\,\mathrm{A}_{t}).
\end{array}
\end{align}
In this section, on the one hand, we show that the map $
\mathrm{\textbf{C}}_{\ell,\texttt{R}}:\Re_\ell(q,s)\rightarrow\texttt{Cy}(\texttt{R},\ell)$
is bijective and the other hand we equip the set
$\mathrm{\textbf{C}}_{\ell,\texttt{R}}$ of binary operations
$\vee$ and  $\wedge$ such that $
\mathrm{\textbf{C}}_{\ell,\texttt{R}}:\left\langle\Re_\ell(q,s)\;;\;\vee,\wedge\right\rangle\rightarrow\left\langle\texttt{Cy}(\texttt{R},\ell)\;;\;+,\cap\,\right\rangle$
is a lattice homomorphism.

 The following theorem gives the number of cyclic codes and free
cyclic codes over finite chain rings.

\begin{Lemma}\cite[Theorem 5.1]{BGG14}\label{ct-cy}
Let $\texttt{R}$ be a finite chain ring of invariants $(q,s).$
Then the number of cyclic $\texttt{R}$-linear codes of length
$\ell,$ is equal to $(s+1)^{|\Sigma_\ell(q)|}.$
\end{Lemma}

\begin{Lemma}\cite[Corollary 11.]{Gra09}\label{n-d} A finite lattice is
distributive if and only if it is isomorphic to $\left\langle
2^{\mathrm{E}};\cup,\cap;\emptyset,\mathrm{E}\right\rangle,$ where
$\mathrm{E}$ is a finite set.
\end{Lemma}

Lemmas\,\ref{ct-cy} and  \ref{n-d} give the following fact.

\begin{Theorem} Let $\texttt{R}$ be a finite chain ring of invariants $(q,s)$ and $\ell$ be a nonnegative integer such that
$\texttt{gcd}(\ell,q)=1.$ Then $s\neq 1$ if and only if
$\left\langle\texttt{Cy}(\texttt{R},\ell)\;;\;+,\cap;
\{\underline{\textbf{0}}\} ,\texttt{R}^\ell\,\right\rangle$ is not
distributive. Moreover, its sublattice
$\left\langle\,\textbf{C}_\ell(\texttt{R});+,\cap;
\{\underline{\textbf{0}}\} ,\texttt{R}^\ell\,\right\rangle$ is
distributive.
\end{Theorem}

 We show that each cyclic $\texttt{R}$-linear code can be written as a direct
sum of irreducibles in precisely one way.

\begin{Lemma}\label{decy} Let $\texttt{R}$ be a finite chain ring of invariants $(q,s).$ Then the
map
$\mathrm{\textbf{C}}_{\ell,\texttt{R}}:\Re_\ell(q,s)\rightarrow
\texttt{Cy}(\texttt{R},\ell)$ is a bijection and the type of $
\mathrm{\textbf{C}}_{\ell,\texttt{R}}(\underline{\mathrm{A}})$ is
$(|\complement_q(\mathrm{A}_0)|,|\complement_q(\mathrm{A}_1)|,\cdots,|\complement_q(\mathrm{A}_{s-1})|),$
for some
$\underline{\mathrm{A}}:=(\mathrm{A}_0,\mathrm{A}_1,\cdots,\mathrm{A}_s)\in\Re_\ell(q,s).$
\end{Lemma}

\begin{Proof} Let $\mathcal{C}$ be an cyclic $\texttt{R}$-linear code of length $\ell.$
From Proposition\,\ref{c(a)}, we have
$$\texttt{R}^\ell=\mathrm{\textbf{C}}_\eta(\texttt{R}\,;\,\Sigma_\ell(q))=\underset{z\in\Sigma_\ell(q)}{\bigoplus}\mathrm{\textbf{C}}_\eta(\texttt{R}\,;\,\{z\})$$
and $\mathrm{\textbf{C}}_\eta(\texttt{R}\,;\,\{z\})$'s are free
irreducible cyclic $\texttt{R}$-linear codes. It follows that
$\mathcal{C}=\underset{z\in\Sigma_\ell(q)}{\bigoplus}\mathcal{C}_z,$
where
$\mathcal{C}_z=\mathrm{\textbf{C}}_\eta(\texttt{R}\,;\,\{z\})\cap\mathcal{C}.$
By
 Proposition\,\ref{irr(a)},
$\mathcal{C}_z=\theta^{t_z}\mathrm{\textbf{C}}_\eta(\texttt{R}\,;\,\{z\}),$
where $t_z\in\{0,1,\cdots,s\}.$ Thus
$\underset{z\in\Sigma_\ell(q)}{\bigoplus}\theta^{t_z}\mathrm{\textbf{C}}_\eta(\texttt{R}\,;\,\{z\})
=\mathrm{\textbf{C}}_{\ell,\texttt{R}}(\mathrm{A}_0,\mathrm{A}_1,\cdots,\mathrm{A}_s),$
where $\mathrm{A}_t=\{z\in\Sigma_\ell\;:\;t_z=t\}.$  Since
$|\Re_\ell(q,s)|=(s+1)^{|\Sigma_\ell(q)|},$ by
Theorem\,\ref{ct-cy}, the uniqueness of
$\underline{\mathrm{A}}:=(\mathrm{A}_0,\mathrm{A}_1,\cdots,\mathrm{A}_s)\in\Re_\ell(q,s)$
such that $\mathcal{C}=
\mathrm{\textbf{C}}_{\ell,\texttt{R}}(\mathrm{A})$ is guaranteed.
Moreover, for every $t\in\{0,1,\cdots,s-1\},$ the cyclic
$\texttt{R}$-linear code
$\mathrm{\textbf{C}}_\eta(\texttt{R}\,;\,\mathrm{A}_{t})$ is free
and
$\texttt{rank}_\texttt{R}(\mathrm{\textbf{C}}_\eta(\texttt{R}\,;\,\mathrm{A}_{t}))=|\complement_q(\mathrm{A}_{t})|.$
Since the direct sum
$\overset{s-1}{\underset{t=0}{\bigoplus}}\theta^{t}\mathrm{\textbf{C}}_\eta(\texttt{R}\,;\,\mathrm{A}_{t})$
gives the type of $
\mathrm{\textbf{C}}_{\ell,\texttt{R}}(\underline{\mathrm{A}}),$
the type of $
\mathrm{\textbf{C}}_{\ell,\texttt{R}}(\underline{\mathrm{A}})$ is
$(k_0,k_1,\cdots,k_{s-1}),$ where
$k_t:=|\complement_q(\mathrm{A}_t)|,$ for every $0\leq t<s-1.$
\end{Proof}

\begin{Definition} Let $\mathcal{C}$ be a cyclic $\texttt{R}$-linear codes of length
$\ell.$ The \emph{defining multiset} of $\mathcal{C}$ is the
$(q,s)$-cyclotomic partition $\underline{\mathrm{A}}$ modulo
$\ell,$ such that $\mathcal{C}=
\mathrm{\textbf{C}}_{\ell,\texttt{R}}(\underline{\mathrm{A}}).$
\end{Definition}

\begin{Proposition}\label{complement} Let
$\underline{\mathrm{A}}:=(\mathrm{A}_0,\mathrm{A}_1,\cdots,\mathrm{A}_s)\in\Re_\ell(q,s)$
and $t\in\{0,1,\cdots,s-1\}.$ Then $
\mathrm{\textbf{C}}_{\ell,\texttt{R}}(\underline{\mathrm{A}})^\perp=
\mathrm{\textbf{C}}_{\ell,\texttt{R}}(\underline{\mathrm{A}}^{\widetilde{\diamond}}),$
where
$\underline{\mathrm{A}}^{\widetilde{\diamond}}:=(-\mathrm{A}_s,-\mathrm{A}_{s-1},\cdots,-\mathrm{A}_1,-\mathrm{A}_0).$
\end{Proposition}

\begin{Proof} Let $\underline{\mathrm{A}}:=(\mathrm{A}_0,\mathrm{A}_1,\cdots,\mathrm{A}_s)\in\Re_\ell(q,s).$ We have
$
\mathrm{\textbf{C}}_{\ell,\texttt{R}}(\underline{\mathrm{A}})^\perp
\supseteq
\bigcap_{u=0}^{s-1}\left(\theta^{s-u}\texttt{R}^\ell+\mathrm{\textbf{C}}_\eta(\texttt{R}\,;\,\mathrm{A}_u^\diamond)\right)$
and
$$\theta^{s-t}\mathrm{\textbf{C}}_\eta(\texttt{R}\,;\,-\mathrm{A}_{t})\subseteq\bigcap_{u=0}^{s-1}\left(\theta^{s-u}\texttt{R}^\ell+\mathrm{\textbf{C}}_\eta(\texttt{R}\,;\,\mathrm{A}_u^\diamond)\right),$$
for every $t\in\{1,2,\cdots,s\}.$  It follows that $
\mathrm{\textbf{C}}_{\ell,\texttt{R}}(-\mathrm{A}_s,-\mathrm{A}_{s-1},\cdots,-\mathrm{A}_1,-\mathrm{A}_0)\subseteq
\mathrm{\textbf{C}}_{\ell,\texttt{R}}(\underline{\mathrm{A}})^\perp.$
From Proposition\,\ref{dual-type} and Theorem\,\ref{main}, $
\mathrm{\textbf{C}}_{\ell,\texttt{R}}(-\mathrm{A}_s,-\mathrm{A}_{s-1},\cdots,-\mathrm{A}_1,-\mathrm{A}_0)$
and $
 \mathrm{\textbf{C}}_{\ell,\texttt{R}}(\underline{\mathrm{A}})^\perp$ have
the same type, we have
$$ \mathrm{\textbf{C}}_{\ell,\texttt{R}}(\underline{\mathrm{A}})^\perp=
 \mathrm{\textbf{C}}_{\ell,\texttt{R}}(-\mathrm{A}_s,-\mathrm{A}_{s-1},\cdots,-\mathrm{A}_1,-\mathrm{A}_0).$$
 \end{Proof}

\begin{Corollary}\label{self-dual} Let
$\underline{\mathrm{A}}:=(\mathrm{A}_0,\mathrm{A}_1,\cdots,\mathrm{A}_s)\in\Re_\ell(q,s).$
Then $
\mathrm{\textbf{C}}_{\ell,\texttt{R}}(\underline{\mathrm{A}})$ is
self-dual if and only if $\mathrm{A}_t=-\mathrm{A}_{s-t},$ for
every $t\in\{0,1\cdots,s\}.$
\end{Corollary}

\begin{Corollary} Let $\texttt{R}$ be a finite chain ring of invariants $(q,s)$ and $s$ is an even integer. Then the following are equivalent.
\begin{enumerate}
    \item there exists a subset $\mathrm{A}$ of $\Sigma_\ell$ such that $\mathrm{A}\neq -\mathrm{A};$
    \item the nontrivial self-dual cyclic $\texttt{R}$-linear codes of length $\ell$ exist.
\end{enumerate}
\end{Corollary}

\begin{Proof} Let $\texttt{R}$ be a finite chain ring of invariants $(q,s)$ and $s$ is an even integer.
\begin{description}
    \item[$1.\Rightarrow 2.$] Assume that there exists a subset $\mathrm{A}$ of $\Sigma_\ell$ such that
$\complement_q(\mathrm{A})\neq -\complement_q(\mathrm{A}).$ Set
$u=\frac{s}{2},$  and
$\mathrm{B}:=\overline{\mathrm{A}\cup(-\mathrm{A})}.$ Consider
$$\mathcal{C}:=\mathrm{\textbf{C}}_{\ell,\texttt{R}}(\cdots,\emptyset,\mathrm{A}_{u-1},\mathrm{A}_u,\mathrm{A}_{u+1},\emptyset,\cdots),$$
where $\mathrm{A}_{u-1}:=\complement_q(\mathrm{A}),$
$\mathrm{A}_{u}:=\complement_q(\mathrm{B})$ and
$\mathrm{A}_{u+1}:=-\mathrm{A}_{u-1}.$ We have
$\mathrm{A}_{s-u+1}=-\mathrm{A}_{u+1},$
$\mathrm{A}_{u}=-\mathrm{A}_{s-u}=-\mathrm{A}_{u}$ and
$\mathrm{A}_t=-\mathrm{A}_{s-t}=\emptyset,$ for every
$t\in\{0,1,\cdots,u-2\}.$ Thanks to Corollary\,\ref{self-dual}, we
can affirm that $\mathcal{C}$ is self-dual.
    \item[$2.\Rightarrow 1.$] Now, we assume that $\mathcal{C}$ is self-dual cyclic
$\texttt{R}$-linear codes of length $\ell$ and every subset
$\mathrm{A}$ of $\Sigma_\ell$ satisfies $\mathrm{A} =-\mathrm{A}.$
Set $\mathcal{C}=
\mathrm{\textbf{C}}_{\ell,\texttt{R}}(\mathrm{A}_0,\mathrm{A}_1,\mathrm{A}_2,\cdots,\mathrm{A}_s).$
From Corollary\,\ref{self-dual}, $\mathrm{A}_t=-\mathrm{A}_{s-t},$
for every $t\in\{0,1,\cdots,s\}.$ Since
$-\mathrm{A}_{s-t}=\mathrm{A}_{s-t},$ we have
$\mathrm{A}_t=\mathrm{A}_{s-t}=\emptyset$ for every
$t\in\{0,1\cdots,s\}\setminus\{\frac{s}{2}\}$ and
$\mathrm{A}_{\frac{s}{2}}=\Sigma_\ell(q).$ Therefore,
$\mathcal{C}=
\mathrm{\textbf{C}}_{\ell,\texttt{R}}(\cdots,\emptyset,\Sigma_\ell(q),\emptyset,\cdots)=\theta^{\frac{s}{2}}\texttt{R}^\ell,$
which is the trivial self-dual code.
\end{description}
\end{Proof}

We point out that the number of cyclic self-dual linear codes over
finite chain rings has been given in \cite{BGG14}. In order to
determine the defining multiset of the sum, and the intersection
of $\texttt{R}$-linear cyclic codes, we extend the binary
operation  $\cup$ of $\Re_\ell(q)$ to $\Re_\ell(q,s)$ as follows:

\begin{Notation}  Let $\underline{\mathrm{A}}:=(\mathrm{A}_0,\mathrm{A}_1,\cdots,\mathrm{A}_s)$
and
$\underline{\mathrm{B}}:=(\mathrm{B}_0,\mathrm{B}_1,\cdots,\mathrm{B}_s)$
be elements of $\;\Re_\ell(q,s).$ Then
\begin{enumerate}
       \item $\underline{\mathrm{A}}\sqcup\underline{\mathrm{B}}:=(\mathrm{C}_0,\mathrm{C}_1,\cdots,\mathrm{C}_s),$ where $\mathrm{C}_0:=\mathrm{A}_0\cup\mathrm{B}_0$
       and $\mathrm{C}_t:=\left(\mathrm{A}_t\cup\mathrm{B}_t\right)\setminus\left(\overset{t-1}{\underset{u=0}{\cup}}\mathrm{C}_u\right),$ for every $0<t\leq s.$
    \item
    $\underline{\mathrm{A}}\sqcap\underline{\mathrm{B}}:=(\underline{\mathrm{A}}^{\widetilde{\diamond}}\vee\underline{\mathrm{B}}^{\widetilde{\diamond}})^{\widetilde{\diamond}}.$
    \end{enumerate}
\end{Notation}

\begin{Theorem}\label{main} The map (\ref{map0}), $ \mathrm{\textbf{C}}_{\ell,\texttt{R}} : \left\langle\,\Re_\ell(q,s); \sqcup, \sqcap; \underline{\emptyset} ,\underline{\Sigma_\ell(q)}\,\right\rangle\rightarrow \left\langle\,\texttt{Cy}(\texttt{R},\ell);+,\,\cap;\,\{\underline{\textbf{0}}\},\,\texttt{R}^\ell\right\rangle$
 is a lattice isomorphism, where
$\underline{\emptyset}:=\left(\emptyset,\cdots,\emptyset,\Sigma_\ell(q)\right)$
and
$\underline{\Sigma_\ell(q)}:=\left(\Sigma_\ell(q),\emptyset,\cdots,\emptyset\right).$
\end{Theorem}

\begin{Proof} Let $\underline{\mathrm{A}}:=(\mathrm{A}_0,\mathrm{A}_1,\cdots,\mathrm{A}_s)\in\Re_\ell(q,s),$
and
$\underline{\mathrm{B}}:=(\mathrm{B}_0,\mathrm{B}_1,\cdots,\mathrm{B}_s)\in\Re_\ell(q,s).$
Firstly, we have
    \begin{eqnarray*}
 \mathrm{\textbf{C}}_{\ell,\texttt{R}}(\underline{\mathrm{A}})+ \mathrm{\textbf{C}}_{\ell,\texttt{R}}(\underline{\mathrm{B}}) &=& \sum_{t=0}^{s-1}\theta^{t}\left(\mathrm{\textbf{C}}_\eta(\texttt{R}\,;\,\mathrm{A}_{t})+\mathrm{\textbf{C}}_\eta(\texttt{R}\,;\,\mathrm{B}_{t})\right), \text{by the associativity of } +, \\
       &=&
       \mathrm{\textbf{C}}_\eta(\texttt{R}\,;\,\mathrm{A}_0\cup\mathrm{B}_0)\oplus\theta\mathrm{\textbf{C}}_\eta(\texttt{R}\,;\,(\mathrm{A}_1\cup\mathrm{B}_1)\setminus\left(\mathrm{A}_0\cup\mathrm{B}_0\right))\oplus\cdots\\
       & & \cdots \oplus\theta^t\mathrm{\textbf{C}}_\eta\left(\texttt{R}\,;\,(\mathrm{A}_t\cup\mathrm{B}_t)\setminus\left(\overset{t-1}{\underset{u=0}{\cup}}(\mathrm{A}_u\cup\mathrm{B}_u)\right)\right)\oplus\cdots\\
       & &
       \cdots\oplus\theta^{s-1}\mathrm{\textbf{C}}_\eta\left(\texttt{R}\,;\,(\mathrm{A}_{s-1}\cup\mathrm{B}_{s-1})\setminus\left(\overset{s-2}{\underset{u=0}{\cup}}(\mathrm{A}_u\cup\mathrm{B}_u)\right)\right)\\
       &=& \mathrm{\textbf{C}}_{\ell,\texttt{R}}(\underline{\mathrm{A}}\sqcup\underline{\mathrm{B}}).
   \end{eqnarray*}
From Propositions\;\ref{dual-ope} and \ref{complement}, we deduce
that
$\mathrm{\textbf{C}}_{\ell,\texttt{R}}(\underline{\mathrm{A}})\cap
\mathrm{\textbf{C}}_{\ell,\texttt{R}}(\underline{\mathrm{B}})=\mathrm{\textbf{C}}_{\ell,\texttt{R}}(\underline{\mathrm{A}}\sqcap\underline{\mathrm{B}}).$
Finally, by Lemma \ref{decy}, we have the expected result.
\end{Proof}

\begin{Proposition}(BCH-bound) Let $\mathcal{C}$ be a cyclic $\texttt{R}$-linear code of length
$\ell$ and
$\underline{\mathrm{A}}:=(\mathrm{A}_0,\mathrm{A}_1,\cdots,\mathrm{A}_s)$
be the element of $\Re_\ell(q,s)$ such that $\mathcal{C}:=
\mathrm{\textbf{C}}_{\ell,\texttt{R}}(\underline{\mathrm{A}})$ and
$\complement_d\left(\overline{\overset{s-1}{\underset{t=0}{\cup}}(-\mathrm{A}_t)}\right)$
contains an interval of length $\delta.$ Then
$$\texttt{Annih}_{\mathcal{C}}(\theta)=\mathrm{\textbf{C}}_{\ell,\texttt{R}}\left(\emptyset,\cdots,\emptyset,\overset{s-1}{\underset{t=0}{\cup
}}\mathrm{A}_t,\mathrm{A}_s\right)\text{ and
}\texttt{wt}(\mathcal{C})\geq 1+\delta.$$
\end{Proposition}

\begin{Proof} By
Theorem\,\ref{dis}, we have $\texttt{Annih}_{\mathcal{C}}(\theta)=
\mathrm{\textbf{C}}_{\ell,\texttt{R}}\left(\emptyset,\cdots,\emptyset,\overset{s-1}{\underset{t=0}{\cup
}}\mathrm{A}_t,\mathrm{A}_s\right),$ and
$\texttt{wt}(\mathcal{C})=\texttt{wt}(\texttt{Annih}_{\mathcal{C}}(\theta)).$
By Theorem \ref{bch}, the lower bound on the minimum Hamming
distance is obtained.
\end{Proof}

 \begin{Example} We again pick $\ell=7,$ $s=2$ and $p=2.$
\begin{center}
\begin{tabular}{cccc}
  \hline
  ~$\mathrm{\textbf{C}}_{\mathbb{Z}_4}\left(\mathrm{A}_0,\mathrm{A}_1,\mathrm{A}_2\right)$~ & ~BHC-bound~     & ~type\;:\;$\left(|\mathrm{A_0}|,|\mathrm{A_1}|\right)$~ & ~Cardinality\;:\;$2^{2|\mathrm{A_0}|+|\mathrm{A_1}|}$  \\
  \hline
  $\mathcal{C}_0:=\mathrm{\textbf{C}}_{\mathbb{Z}_4}\left(\emptyset,\emptyset,\complement_2(\{0,1,3\})\right)$ & $0$  & (0,0) &   1 \\
  $\mathcal{C}_{26}:=\mathrm{\textbf{C}}_{\mathbb{Z}_4}\left(\complement_2(\{0,1,3\}),\emptyset,\emptyset\right)$ & $1$  & (7,0) &  $2^{14}$ \\
  $\mathcal{C}_{23}:=\mathrm{\textbf{C}}_{\mathbb{Z}_4}\left(\complement_2(\{1,3\}),\complement_2(\{0\}),\emptyset\right)$ & $1$ &  (6,0) &  $2^{12}$ \\
  $\mathcal{C}_{22}:=\mathrm{\textbf{C}}_{\mathbb{Z}_4}\left(\complement_2(\{0,3\}),\complement_2(\{1\}),\emptyset\right)$ & $1$ &  (4,3) &  $2^{11}$ \\
  $\mathcal{C}_{24}:=\mathrm{\textbf{C}}_{\mathbb{Z}_4}\left(\complement_2(\{0,1\}),\complement_2(\{3\}),\emptyset\right)$ & $1$ &  (4,3) &   $2^{11}$ \\
  $\mathcal{C}_{21}:=\mathrm{\textbf{C}}_{\mathbb{Z}_4}\left(\complement_2(\{3\}),\complement_2(\{0,1\}),\emptyset\right)$ & $1$ &  (3,4) &   $2^{10}$ \\
  $\mathcal{C}_{20}:=\mathrm{\textbf{C}}_{\mathbb{Z}_4}\left(\complement_2(\{1\}),\complement_2(\{0,3\}),\emptyset\right)$ & $1$ &   (3,4) &  $2^{10}$ \\
  $\mathcal{C}_{17}:=\mathrm{\textbf{C}}_{\mathbb{Z}_4}\left(\complement_2(\{0\}),\complement_2(\{1,3\}),\emptyset\right)$ & $1$ &  (1,6) &  $2^8$ \\
  $\mathcal{C}_{14}:=\mathrm{\textbf{C}}_{\mathbb{Z}_4}\left(\emptyset,\complement_2(\{0,1,3\}),\emptyset\right)$ & $1$  & (0,7) &  $2^7$ \\
  $\mathcal{C}_{25}:=\mathrm{\textbf{C}}_{\mathbb{Z}_4}\left(\complement_2(\{1,3\}),\emptyset,\complement_2(\{0\})\right)$ & $2$ &  (6,0) &  $2^{12}$ \\
  $\mathcal{C}_{19}:=\mathrm{\textbf{C}}_{\mathbb{Z}_4}\left(\complement_2(\{1\}),\complement_2(\{3\}),\complement_2(\{0\})\right)$  & $2$ &  (3,3) &  $2^9$ \\
  $\mathcal{C}_{18}:=\mathrm{\textbf{C}}_{\mathbb{Z}_4}\left(\complement_2(\{3\}),\complement_2(\{1\}),\complement_2(\{0\})\right)$ & $2$ &  (3,3) &  $2^9$ \\
  $\mathcal{C}_9:=\mathrm{\textbf{C}}_{\mathbb{Z}_4}\left(\emptyset,\complement_2(\{1,3\}),\complement_2(\{0\})\right)$ & $2$ &  (0,6) &  $2^6$ \\
  $\mathcal{C}_{16}:=\mathrm{\textbf{C}}_{\mathbb{Z}_4}\left(\complement_2(\{0,3\}),\emptyset,\complement_2(\{1\})\right)$ & $3$ & (4,0) &  $2^8$ \\
  $\mathcal{C}_{13}:=\mathrm{\textbf{C}}_{\mathbb{Z}_4}\left(\complement_2(\{3\}),\complement_2(\{0\}),\complement_2(\{1\})\right)$ & $3$ &  (3,1) &  $2^{7}$ \\
  $\mathcal{C}_{12}:=\mathrm{\textbf{C}}_{\mathbb{Z}_4}\left(\complement_2(\{1\}),\complement_2(\{0\}),\complement_2(\{3\})\right)$ & $3$ &  (3,1) &   $2^7$ \\
  $\mathcal{C}_8:=\mathrm{\textbf{C}}_{\mathbb{Z}_4}\left(\complement_2(\{0\}),\complement_2(\{1\}),\complement_2(\{3\})\right)$ & $3$ &  (1,3) &   $2^5$ \\
  $\mathcal{C}_7:=\mathrm{\textbf{C}}_{\mathbb{Z}_4}\left(\complement_2(\{0\}),\complement_2(\{3\}),\complement_2(\{1\})\right)$ & $3$ &  (1,3) &   $2^5$ \\
  $\mathcal{C}_5:=\mathrm{\textbf{C}}_{\mathbb{Z}_4}\left(\emptyset,\complement_2(\{0,3\}),\complement_2(\{1\})\right)$ & $3$ & (0,4) &   $2^{4}$ \\
  $\mathcal{C}_{11}:=\mathrm{\textbf{C}}_{\mathbb{Z}_4}\left(\complement_2(\{1\}),\emptyset,\complement_2(\{0,3\})\right)$ & $4$ &  (3,0) &   $2^6$ \\
  $\mathcal{C}_{10}:=\mathrm{\textbf{C}}_{\mathbb{Z}_4}\left(\complement_2(\{3\}),\emptyset,\complement_2(\{0,1\})\right)$ & $4$ &  (3,0) &  $2^{6}$ \\
  $\mathcal{C}_4:=\mathrm{\textbf{C}}_{\mathbb{Z}_4}\left(\emptyset,\complement_2(\{1\}),\complement_2(\{0,3\})\right)$ & $4$ &  (0,3) &  $2^3$ \\
  $\mathcal{C}_3:=\mathrm{\textbf{C}}_{\mathbb{Z}_4}\left(\emptyset,\complement_2(\{3\}),\complement_2(\{0,1\})\right)$ & $4$ &  (0,3) &  $2^3$ \\
  $\mathcal{C}_{15}:=\mathrm{\textbf{C}}_{\mathbb{Z}_4}\left(\complement_2(\{0,1\}),\emptyset,\complement_2(\{3\})\right)$ & $4$ &  (4,0) &   $2^{8}$ \\
  $\mathcal{C}_6:=\mathrm{\textbf{C}}_{\mathbb{Z}_4}\left(\emptyset,\complement_2(\{0,1\}),\complement_2(\{3\})\right)$ & $4$ & (0,4) &   $2^4$ \\
  $\mathcal{C}_2:=\mathrm{\textbf{C}}_{\mathbb{Z}_4}\left(\complement_2(\{0\}),\emptyset,\complement_2(\{1,3\})\right)$ & $7$ &(1,0) &   $2^2$ \\
 $\mathcal{C}_1:=\mathrm{\textbf{C}}_{\mathbb{Z}_4}\left(\emptyset,\complement_2(\{0\}),\complement_2(\{1,3\})\right)$ & $7$ &   (0,1) &  $2^1$ \\
  \hline
\end{tabular}
\captionof{table}{Cyclic $\mathbb{Z}_4$-linear codes of length
$7.$}
\end{center}

We have
$\mathcal{C}_{8}+\mathcal{C}_{12}=\mathcal{C}_{15},\mathcal{C}_{19}+\mathcal{C}_{12}=\mathcal{C}_{20},$
$\mathcal{C}_{8}^\perp=\mathcal{C}_{19}$ and
$\mathcal{C}_{12}^\perp=\mathcal{C}_{12}.$ So
$\mathcal{C}_{8}\cap\mathcal{C}_{12}=(\mathcal{C}_{8}^\perp+\mathcal{C}_{12}^\perp)^\perp=(\mathcal{C}_{19}+\mathcal{C}_{12})^\perp=\mathcal{C}_{20}^\perp
= \mathcal{C}_{6}.$

\end{Example}

\end{document}